%
%
%
%
\def\unredoffs{} \def\redoffs{\voffset=-.40truein\hoffset=-.40truein}
\def\speclscape{}
%
%
%
%
\newbox\leftpage \newdimen\fullhsize \newdimen\hstitle \newdimen\hsbody
\tolerance=1000\hfuzz=2pt
\catcode`\@=11 
\def\bigans{b }
\def\answ{b }
\ifx\answ\bigans\message{(This will come out unreduced.}
\magnification=1200\unredoffs\baselineskip=16pt plus 2pt minus 1pt
\hsbody=\hsize \hstitle=\hsize 
\else\message{(This will be reduced.} \let\l@r=L
\magnification=1000\baselineskip=16pt plus 2pt minus 1pt \vsize=7truein
\redoffs \hstitle=8truein\hsbody=4.75truein\fullhsize=10truein\hsize=\hsbody
\output={\ifnum\pageno=0 
   \shipout\vbox{\speclscape{\hsize\fullhsize\makeheadline}
     \hbox to \fullhsize{\hfill\pagebody\hfill}}\advancepageno
   \else
   \almostshipout{\leftline{\vbox{\pagebody\makefootline}}}\advancepageno
   \fi}
\def\almostshipout#1{\if L\l@r \count1=1 \message{[\the\count0.\the\count1]}
       \global\setbox\leftpage=#1 \global\let\l@r=R
  \else \count1=2
   \shipout\vbox{\speclscape{\hsize\fullhsize\makeheadline}
       \hbox to\fullhsize{\box\leftpage\hfil#1}}  \global\let\l@r=L\fi}
\fi
%
\newcount\yearltd\yearltd=\year\advance\yearltd by -1900

\def\Title#1#2{\nopagenumbers\abstractfont\hsize=\hstitle\rightline{#1}%
\vskip 1in\centerline{\titlefont #2}\abstractfont\vskip .5in\pageno=0}
\def\Date#1{\vfill\leftline{#1}\tenpoint\supereject\global\hsize=\hsbody%
\footline={\hss\tenrm\folio\hss}}
%

\def\draftmode{\message{ DRAFTMODE }\def\draftdate{{\rm preliminary draft:
\number\month/\number\day/\number\yearltd\ \ \hourmin}}%
\headline={\hfil\draftdate}\writelabels\baselineskip=20pt plus 2pt minus 2pt
  {\count255=\time\divide\count255 by 60 \xdef\hourmin{\number\count255}
   \multiply\count255 by-60\advance\count255 by\time
   \xdef\hourmin{\hourmin:\ifnum\count255<10 0\fi\the\count255}}}
\def\nolabels{\def\wrlabeL##1{}\def\eqlabeL##1{}\def\reflabeL##1{}}
\def\writelabels{\def\wrlabeL##1{\leavevmode\vadjust{\rlap{\smash%
{\line{{\escapechar=` \hfill\rlap{\sevenrm\hskip.03in\string##1}}}}}}}%
\def\eqlabeL##1{{\escapechar-1\rlap{\sevenrm\hskip.05in\string##1}}}%
\def\reflabeL##1{\noexpand\llap{\noexpand\sevenrm\string\string\string##1}}}
\nolabels
%
\global\newcount\secno \global\secno=0
\global\newcount\meqno \global\meqno=1
\def\newsec#1{\global\advance\secno by1\message{(\the\secno. #1)}
\global\subsecno=0\eqnres@t\noindent{\bf\the\secno. #1}
\writetoca{{\secsym} {#1}}\par\nobreak\medskip\nobreak}
\def\eqnres@t{\xdef\secsym{\the\secno.}\global\meqno=1\bigbreak\bigskip}
\def\sequentialequations{\def\eqnres@t{\bigbreak}}\xdef\secsym{}
\global\newcount\subsecno \global\subsecno=0
\def\subsec#1{\global\advance\subsecno by1\message{(\secsym\the\subsecno. #1)}
\ifnum\lastpenalty>9000\else\bigbreak\fi
\noindent{\it\secsym\the\subsecno. #1}\writetoca{\string\quad
{\secsym\the\subsecno.} {#1}}\par\nobreak\medskip\nobreak}
\def\appendix#1#2{\global\meqno=1\global\subsecno=0\xdef\secsym{\hbox{#1.}}
\bigbreak\bigskip\noindent{\bf Appendix #1. #2}\message{(#1. #2)}
\writetoca{Appendix {#1.} {#2}}\par\nobreak\medskip\nobreak}
%
%
\def\eqnn#1{\xdef #1{(\secsym\the\meqno)}\writedef{#1\leftbracket#1}%
\global\advance\meqno by1\wrlabeL#1}
\def\eqna#1{\xdef #1##1{\hbox{$(\secsym\the\meqno##1)$}}
\writedef{#1\numbersign1\leftbracket#1{\numbersign1}}%
\global\advance\meqno by1\wrlabeL{#1$\{\}$}}
\def\eqn#1#2{\xdef #1{(\secsym\the\meqno)}\writedef{#1\leftbracket#1}%
\global\advance\meqno by1$$#2\eqno#1\eqlabeL#1$$}
%
\newskip\footskip\footskip14pt plus 1pt minus 1pt 
\def\footnotefont{\ninepoint}\def\f@t#1{\footnotefont #1\@foot}
\def\f@@t{\baselineskip\footskip\bgroup\footnotefont\aftergroup\@foot\let\next}
\setbox\strutbox=\hbox{\vrule height9.5pt depth4.5pt width0pt}
\global\newcount\ftno \global\ftno=0
\def\foot{\global\advance\ftno by1\footnote{$^{\the\ftno}$}}
%
\newwrite\ftfile
\def\footend{\def\foot{\global\advance\ftno by1\chardef\wfile=\ftfile
$^{\the\ftno}$\ifnum\ftno=1\immediate\openout\ftfile=foots.tmp\fi%
\immediate\write\ftfile{\noexpand\smallskip%
\noexpand\item{f\the\ftno:\ }\pctsign}\findarg}%
\def\footatend{\vfill\eject\immediate\closeout\ftfile{\parindent=20pt
\centerline{\bf Footnotes}\nobreak\bigskip\input foots.tmp }}}
\def\footatend{}
%
%
\global\newcount\refno \global\refno=1
\newwrite\rfile
\def\ref{[\the\refno]\nref}
\def\nref#1{\xdef#1{[\the\refno]}\writedef{#1\leftbracket#1}%
\ifnum\refno=1\immediate\openout\rfile=refs.tmp\fi
\global\advance\refno by1\chardef\wfile=\rfile\immediate
\write\rfile{\noexpand\item{#1\ }\reflabeL{#1\hskip.31in}\pctsign}\findarg}
\def\findarg#1#{\begingroup\obeylines\newlinechar=`\^^M\pass@rg}
{\obeylines\gdef\pass@rg#1{\writ@line\relax #1^^M\hbox{}^^M}%
\gdef\writ@line#1^^M{\expandafter\toks0\expandafter{\striprel@x #1}%
\edef\next{\the\toks0}\ifx\next\em@rk\let\next=\endgroup\else\ifx\next\empty%
\else\immediate\write\wfile{\the\toks0}\fi\let\next=\writ@line\fi\next\relax}}
\def\striprel@x#1{} \def\em@rk{\hbox{}}
\def\lref{\begingroup\obeylines\lr@f}
\def\lr@f#1#2{\gdef#1{\ref#1{#2}}\endgroup\unskip}
\def\semi{;\hfil\break}
\def\addref#1{\immediate\write\rfile{\noexpand\item{}#1}} 
\def\startrefs#1{\immediate\openout\rfile=refs.tmp\refno=#1}
\def\xref{\expandafter\xr@f}\def\xr@f[#1]{#1}
\def\refs#1{\count255=1[\r@fs #1{\hbox{}}]}
\def\r@fs#1{\ifx\und@fined#1\message{reflabel \string#1 is undefined.}%
\nref#1{need to supply reference \string#1.}\fi%
\vphantom{\hphantom{#1}}\edef\next{#1}\ifx\next\em@rk\def\next{}%
\else\ifx\next#1\ifodd\count255\relax\xref#1\count255=0\fi%
\else#1\count255=1\fi\let\next=\r@fs\fi\next}
%

%
\newwrite\ffile\global\newcount\figno \global\figno=1
\def\fig{fig.~\the\figno\nfig}
\def\nfig#1{\xdef#1{fig.~\the\figno}%
\writedef{#1\leftbracket fig.\noexpand~\the\figno}%
\ifnum\figno=1\immediate\openout\ffile=figs.tmp\fi\chardef\wfile=\ffile%
\immediate\write\ffile{\noexpand\medskip\noexpand\item{Fig.\ \the\figno. }
\reflabeL{#1\hskip.55in}\pctsign}\global\advance\figno by1\findarg}
\def\xfig{\expandafter\xf@g}\def\xf@g fig.\penalty\@M\ {}
\def\figs#1{figs.~\f@gs #1{\hbox{}}}
\def\f@gs#1{\edef\next{#1}\ifx\next\em@rk\def\next{}\else
\ifx\next#1\xfig #1\else#1\fi\let\next=\f@gs\fi\next}
\newwrite\lfile
{\escapechar-1\xdef\pctsign{\string\%}\xdef\leftbracket{\string\{}
\xdef\rightbracket{\string\}}\xdef\numbersign{\string\#}}

\def\writestop{\def\writestoppt{\immediate\write\lfile{\string\pageno%
\the\pageno\string\startrefs\leftbracket\the\refno\rightbracket%
\string\def\string\secsym\leftbracket\secsym\rightbracket%
\string\secno\the\secno\string\meqno\the\meqno}\immediate\closeout\lfile}}
\def\writestoppt{}\def\writedef#1{}
\def\seclab#1{\xdef #1{\the\secno}\writedef{#1\leftbracket#1}\wrlabeL{#1=#1}}
\def\subseclab#1{\xdef #1{\secsym\the\subsecno}%
\writedef{#1\leftbracket#1}\wrlabeL{#1=#1}}
\newwrite\tfile \def\writetoca#1{}
\def\leaderfill{\leaders\hbox to 1em{\hss.\hss}\hfill}
\def\writetoc{\immediate\openout\tfile=toc.tmp
    \def\writetoca##1{{\edef\next{\write\tfile{\noindent ##1
    \string\leaderfill {\noexpand\number\pageno} \par}}\next}}}

%
\catcode`\@=12 
%
\edef\tfontsize{\ifx\answ\bigans scaled\magstep3\else scaled\magstep4\fi}
\font\titlerm=cmr10 \tfontsize \font\titlerms=cmr7 \tfontsize
\font\titlermss=cmr5 \tfontsize \font\titlei=cmmi10 \tfontsize
\font\titleis=cmmi7 \tfontsize \font\titleiss=cmmi5 \tfontsize
\font\titlesy=cmsy10 \tfontsize \font\titlesys=cmsy7 \tfontsize
\font\titlesyss=cmsy5 \tfontsize \font\titleit=cmti10 \tfontsize
\skewchar\titlei='177 \skewchar\titleis='177 \skewchar\titleiss='177
\skewchar\titlesy='60 \skewchar\titlesys='60 \skewchar\titlesyss='60
\def\titlefont{\def\rm{\fam0\titlerm}
\textfont0=\titlerm \scriptfont0=\titlerms \scriptscriptfont0=\titlermss
\textfont1=\titlei \scriptfont1=\titleis \scriptscriptfont1=\titleiss
\textfont2=\titlesy \scriptfont2=\titlesys \scriptscriptfont2=\titlesyss
\textfont\itfam=\titleit \def\it{\fam\itfam\titleit}\rm}
 \ifx\answ\bigans\else scaled\magstep1\fi
\ifx\answ\bigans\def\abstractfont{\tenpoint}\else
\font\abssl=cmsl10 scaled \magstep1
\font\absrm=cmr10 scaled\magstep1 \font\absrms=cmr7 scaled\magstep1
\font\absrmss=cmr5 scaled\magstep1 \font\absi=cmmi10 scaled\magstep1
\font\absis=cmmi7 scaled\magstep1 \font\absiss=cmmi5 scaled\magstep1
\font\abssy=cmsy10 scaled\magstep1 \font\abssys=cmsy7 scaled\magstep1
\font\abssyss=cmsy5 scaled\magstep1 \font\absbf=cmbx10 scaled\magstep1
\skewchar\absi='177 \skewchar\absis='177 \skewchar\absiss='177
\skewchar\abssy='60 \skewchar\abssys='60 \skewchar\abssyss='60
\def\abstractfont{\def\rm{\fam0\absrm}
\textfont0=\absrm \scriptfont0=\absrms \scriptscriptfont0=\absrmss
\textfont1=\absi \scriptfont1=\absis \scriptscriptfont1=\absiss
\textfont2=\abssy \scriptfont2=\abssys \scriptscriptfont2=\abssyss
\textfont\itfam=\bigit \def\it{\fam\itfam\bigit}\def\footnotefont{\tenpoint}%
\textfont\slfam=\abssl \def\sl{\fam\slfam\abssl}%
\textfont\bffam=\absbf \def\bf{\fam\bffam\absbf}\rm}\fi
\def\tenpoint{\def\rm{\fam0\tenrm}
\textfont0=\tenrm \scriptfont0=\sevenrm \scriptscriptfont0=\fiverm
\textfont1=\teni  \scriptfont1=\seveni  \scriptscriptfont1=\fivei
\textfont2=\tensy \scriptfont2=\sevensy \scriptscriptfont2=\fivesy
\textfont\itfam=\tenit \def\it{\fam\itfam\tenit}\def\footnotefont{\ninepoint}%
\textfont\bffam=\tenbf \def\bf{\fam\bffam\tenbf}\def\sl{\fam\slfam\tensl}\rm}
\font\ninerm=cmr9 \font\sixrm=cmr6 \font\ninei=cmmi9 \font\sixi=cmmi6
\font\ninesy=cmsy9 \font\sixsy=cmsy6 \font\ninebf=cmbx9
\font\nineit=cmti9 \font\ninesl=cmsl9 \skewchar\ninei='177
\skewchar\sixi='177 \skewchar\ninesy='60 \skewchar\sixsy='60
\def\ninepoint{\def\rm{\fam0\ninerm}
\textfont0=\ninerm \scriptfont0=\sixrm \scriptscriptfont0=\fiverm
\textfont1=\ninei \scriptfont1=\sixi \scriptscriptfont1=\fivei
\textfont2=\ninesy \scriptfont2=\sixsy \scriptscriptfont2=\fivesy
\textfont\itfam=\ninei \def\it{\fam\itfam\nineit}\def\sl{\fam\slfam\ninesl}%
\textfont\bffam=\ninebf \def\bf{\fam\bffam\ninebf}\rm}
%
%

\hyphenation{anom-aly anom-alies coun-ter-term coun-ter-terms}
\def\inv{^{\raise.15ex\hbox{${\scriptscriptstyle -}$}\kern-.05em 1}}

\def\Dsl{\,\raise.15ex\hbox{/}\mkern-13.5mu D} 
\def\dsl{\raise.15ex\hbox{/}\kern-.57em\partial}

\def\tr{{\rm tr}} \def\Tr{{\rm Tr}}
\font\bigit=cmti10 scaled \magstep1
\def\lspace{\ifx\answ\bigans{}\else\qquad\fi}
\def\lbspace{\ifx\answ\bigans{}\else\hskip-.2in\fi} 
\def\boxeqn#1{\vcenter{\vbox{\hrule\hbox{\vrule\kern3pt\vbox{\kern3pt
     \hbox{${\displaystyle #1}$}\kern3pt}\kern3pt\vrule}\hrule}}}
\def\mbox#1#2{\vcenter{\hrule \hbox{\vrule height#2in
         \kern#1in \vrule} \hrule}}  
%
 \def\CO{{\cal O}} 
\def\CA{{\cal A}}  \def\CF{{\cal F}} 
 \def\CH{{\cal H}}  \def\CU{{\cal U}}
  \def\CD{{\cal D}} 
\def\e#1{{\rm e}^{^{\textstyle#1}}}

\def\darr#1{\raise1.5ex\hbox{$\leftrightarrow$}\mkern-16.5mu #1}

\def\half{{\textstyle{1\over2}}} 
\def\roughly#1{\raise.3ex\hbox{$#1$\kern-.75em\lower1ex\hbox{$\sim$}}}

\def\np#1#2#3{Nucl. Phys. {\bf B#1} (#2) #3}
\def\pl#1#2#3{Phys. Lett. {\bf #1B} (#2) #3}

\def\anp#1#2#3{Ann. Phys. {\bf #1} (#2) #3}

\def\prep#1#2#3{Phys. Rep. {\bf #1} (#2) #3}

\def\cmp#1#2#3{Comm. Math. Phys. {\bf #1} (#2) #3}

\def\jhep#1#2#3{JHEP {\bf#1}(#2) #3}

\def\IB{\relax\hbox{$\inbar\kern-.3em{\rm B}$}}
\def\IC{\relax\hbox{$\inbar\kern-.3em{\rm C}$}}
\def\ID{\relax\hbox{$\inbar\kern-.3em{\rm D}$}}
\def\IE{\relax\hbox{$\inbar\kern-.3em{\rm E}$}}
\def\IF{\relax\hbox{$\inbar\kern-.3em{\rm F}$}}
\def\IG{\relax\hbox{$\inbar\kern-.3em{\rm G}$}}
\def\IGa{\relax\hbox{${\rm I}\kern-.18em\Gamma$}}
\def\IH{\relax{\rm I\kern-.18em H}}
\def\IK{\relax{\rm I\kern-.18em K}}
\def\IL{\relax{\rm I\kern-.18em L}}
\def\IP{\relax{\rm I\kern-.18em P}}
\def\IR{\relax{\rm I\kern-.18em R}}
\def\IZ{\relax\ifmmode\mathchoice{
\hbox{\cmss Z\kern-.4em Z}}{\hbox{\cmss Z\kern-.4em Z}}
{\lower.9pt\hbox{\cmsss Z\kern-.4em Z}}
{\lower1.2pt\hbox{\cmsss Z\kern-.4em Z}}
\else{\cmss Z\kern-.4em Z}\fi}
\def\II{\relax{\rm I\kern-.18em I}}

\def\ndt{{\noindent}}


\def\CA{{\cal A}}

\def\CD{{\cal D}}

\def\CF{{\cal F}}

\def\CH{{\cal H}}

\def\CM{{\cal M}}
\def\CN{{\cal N}}
\def\CO{{\cal O}}
\def\CP{{\cal P}}

\def\CS{{\cal S}}

\def\CU{{\cal U}}

\def\CZ{{\cal Z}}

\def\p{\partial}
\def\pb{\bar{\partial}}



\def\Tr{{\rm Tr}}


\def\inbar{\,\vrule height1.5ex width.4pt depth0pt}

\font\cmss=cmss10 \font\cmsss=cmss10 at 7pt

\def\a{{\alpha}}

\def\b{{\beta}}
\def\d{{\delta}}

\def\e{{\epsilon}}
\def\z{{\zeta}}
\def\ve{{\varepsilon}}
\def\vf{{\varphi}}
\def\m{{\mu}}
\def\n{{\nu}}
\def\u{{\Upsilon}}
\def\l{{\lambda}}
\def\s{{\sigma}}
\def\t{{\theta}}

\def\o{{\omega}}


\def\IF{{\bf F}}
\def\boxit#1{\vbox{\hrule\hbox{\vrule\kern8pt
\vbox{\hbox{\kern8pt}\hbox{\vbox{#1}}\hbox{\kern8pt}}
\kern8pt\vrule}\hrule}}
\def\mathboxit#1{\vbox{\hrule\hbox{\vrule\kern8pt\vbox{\kern8pt
\hbox{$\displaystyle #1$}\kern8pt}\kern8pt\vrule}\hrule}}

\chardef\tempcat=\the\catcode`\@ \catcode`\@=11
\def\cyracc{\def\u##1{\if \i##1\accent"24 i%
    \else \accent"24 ##1\fi }}
\newfam\cyrfam
\font\tencyr=wncyr10
\def\cyr{\fam\cyrfam\tencyr\cyracc}


\def\lref{\begingroup\obeylines\lr@f}
\def\lr@f#1#2{\gdef#1{\ref#1{#2}}\endgroup\unskip}

\lref\torusaction{G.~Ellingsrud, S.A.Stromme, Invent. Math. {\bf
87} (1987) 343-352\semi A.~Klyachko, Math. USSR Izv {\bf 53}
(1989) $n^0$ 5, 1001-1021\semi  A.~Gorodentsev, M.~Leenson,
alg-geom/9604011\semi L.~G\"ottche, Math. Ann. {\bf 286}(1990)
193-207}

\lref\ikkt{N.~Ishibashi, H.~Kawai, Y.~Kitazawa, and A.~Tsuchiya,
\np{498}{1997}{467}, hep-th/9612115}

\lref\cds{A.~Connes, M.~Douglas, A.~Schwarz,
\jhep{9802}{1998}{003}}

\lref\wtnc{E.~Witten, \np{268}{1986}{253}}

\lref\gopakumarvafa{R.~Gopakumar, C.Vafa, hep-th/9809187,
hep-th/9812127}

\lref\wittenone{E.~Witten, hep-th/9403195}

\lref\cg{E.~Corrigan, P.~Goddard, ``Construction of instanton and
monopole solutions and reciprocity'', \anp {154}{1984}{253}}

\lref\opennc{N.~Nekrasov, hep-th/0203109\semi K.-Y.Kim, B.-H. Lee,
H.S. Yang, hep-th/0205010 }

\lref\donaldson{S.K.~Donaldson, ``Instantons and Geometric
Invariant Theory", \cmp{93}{1984}{453-460}}

\lref\nakajima{H.~Nakajima, ``Lectures on Hilbert Schemes of
Points on Surfaces''\semi AMS University Lecture Series, 1999,
ISBN 0-8218-1956-9. }

\lref\neksch{N.~Nekrasov, A.~S.~Schwarz, hep-th/9802068,
\cmp{198}{1998}{689}}

\lref\freck{A.~Losev, N.~Nekrasov, S.~Shatashvili, hep-th/9908204,
hep-th/9911099}

\lref\rkh{N.J.~Hitchin, A.~Karlhede, U.~Lindstrom, and M.~Rocek,
\cmp{108}{1987}{535}}

\lref\branek{H.~Braden, N.~Nekrasov, hep-th/9912019\semi
K.~Furuuchi, hep-th/9912047}

\lref\wilson{G.~ Wilson, ``Collisions of Calogero-Moser particles
and adelic Grassmannian", Invent. Math. 133 (1998) 1-41.}

\lref\avatars{A.~Losev, G.~Moore, N.~Nekrasov, S.~Shatashvili,
hep-th/9509151}

\lref\abkss{O.~Aharony, M.~Berkooz, S.~Kachru, N.~Seiberg,
E.~Silverstein, hep-th/9707079\semi O.~Aharony, M.~Berkooz,
N.~Seiberg, hep-th/9712117\semi R.~Dijkgraaf, hep-th/9810157}

\lref\witsei{N.~Seiberg, E.~Witten, hep-th/9908142,
\jhep{9909}{1999}{032}}

\lref\kkn{V.~Kazakov, I.~Kostov, N.~Nekrasov, ``D-particles,
Matrix Integrals and KP hierarchy'', \np{557}{1999}{413-442},
hep-th/9810035}

\lref\niemi{E.~Witten, Jour. Diff. Geom. {\bf 17} (1982) 661-692
\semi L.~Alvarez-Gaum\'e, \cmp{90}{1983}{161-173}\semi P.~Windey,
Acta~Phys.~Polon. {\bf B15} (1984) 435,M.~Atiyah, Ast\'erisque
{\bf 2} (1985) 43-60 \semi A.~Morozov, A.~Niemi, K.~Palo,
\np{377}{1992}{295-338}\semi L.~Baulieu, A.~Losev, N.~Nekrasov,
hep-th/9707174} \lref\DHf{J.~J.~Duistermaat, G.J.~Heckman, Invent.
Math. {\bf 69} (1982) 259\semi M.~Atiyah, R.~Bott, Topology {\bf
23} No 1 (1984) 1} \lref\tdgt{M.~Atiyah, R.~Bott, Phil. Trans.
Roy. Soc. London {\bf A 308} (1982), 524-615\semi
 E.~Witten, hep-th/9204083\semi S.~Cordes,
G.~Moore, S.~Rangoolam, hep-th/9411210}

\lref\atiyahsegal{M.~Atiyah, G.~Segal, Ann. of Math. {\bf 87}
(1968) 531}

\lref\bott{R.~Bott, J.~Diff.~Geom. {\bf 4} (1967) 311}

\lref\gravilit{M.~Bershadsky, S.~Cecotti, H.~Ooguri, C.~Vafa,
\cmp{165}{1994}{311}, \np{405}{1993}{279}\semi I.~Antoniadis,
E.~Gava, K.S.~Narain, T.~R.~Taylor, \np{413}{1994}{162},
\np{455}{1995}{109}}

\lref\calculus{N.~Dorey, T.~J.~Hollowood, V.~V.~Khoze,
M.~P.~Mattis, hep-th/0206063}

\lref\instmeasures{N.~Dorey, V.V.~Khoze, M.P.~Mattis,
hep-th/9706007, hep-th/9708036}

\lref\twoinst{N.~Dorey, V.V.~Khoze, M.P.~Mattis, hep-th/9607066}

\lref\vafaengine{S.~Katz, A.~Klemm, C.~Vafa, hep-th/9609239}

\lref\connes{A.~Connes, ``Noncommutative geometry'', Academic
Press (1994)}

\lref\macdonald{I.~Macdonald, ``Symmetric functions and Hall
polynomials'', Clarendon Press, Oxford, 1979}

\lref\nikfive{N.~Nekrasov, hep-th/9609219 \semi A.~Lawrence,
N.~Nekrasov, hep-th/9706025}

\lref\instlit{Literature on instantons}

\lref\givental{A.~Givental, alg-geom/9603021}

\lref\maxim{M.~Kontsevich, hep-th/9405035}

\lref\whitham{A.~Gorsky, A.~Marshakov, A.~Mironov, A.~Morozov,
hep-th/9802007}

\lref\kricheverwhitham{I.~Krichever, hep-th/9205110,
\cmp{143}{1992}{415}}

\lref\sw{N.~Seiberg, E.~Witten, hep-th/9407087, hep-th/9408099}

\lref\swsol{A.~Klemm, W.~Lerche, S.~Theisen, S.~Yankielowisz,
hep-th/9411048 \semi P.~Argyres, A.~Faraggi, hep-th/9411057\semi
A.~Hannany, Y.~Oz, hep-th/9505074}

\lref\hollowood{T.~Hollowood, hep-th/0201075, hep-th/0202197}

\lref\nsvz{V.~Novikov, M.~Shifman, A.~Vainshtein, V.~Zakharov,
\pl{217}{1989}{103}}

\lref\seibergone{N.~Seiberg, \pl{206}{1988}{75}}

\lref\dbound{G.~Moore, N.~Nekrasov, S.~Shatashvili,
hep-th/9803265}

\lref\ihiggs{G.~Moore, N.~Nekrasov, S.~Shatashvili, hep-th/9712241
}

\lref\potsdam{W.~Krauth, H.~Nicolai, M.~Staudacher,
hep-th/9803117}

\lref\kirwan{F.~Kirwan, ``Cohomology of quotients in symplectic
and algebraic geometry'', Mathematical Notes, Princeton Univ.
Press, 1985}

\lref\wittfivebrane{E.~Witten, hep-th/9610234}

\lref\issues{A.~Losev, N.~Nekrasov, S.~Shatashvili,
hep-th/9711108, hep-th/9801061}

\lref\adhm{M.~Atiyah, V.~Drinfeld, N.~Hitchin, Yu.~Manin, Phys.
Lett. {\bf 65A} (1978) 185}

\lref\warner{A.~Klemm, W.~Lerche, P.~Mayr, C.~Vafa, N.~Warner,
hep-th/9604034}

\lref\wittensolution{E.~Witten, hep-th/9703166}

\lref\twists{E.~Witten, hep-th/9304026 \semi O.~Ganor,
hep-th/9903110 \semi H.~Braden, A.~Marshakov, A.~Mironov,
A.~Morozov, hep-th/9812078}

\lref\experiment{G.~Chan, E.~D'Hoker, hep-th/9906193 \semi
E.~D'Hoker, I.~Krichever, D.~Phong, hep-th/9609041\semi
J.~Edelstein, M.~Gomez-Reino, J.~Mas, hep-th/9904087 \semi
J.~Edelstein, M.~Mari\~no, J.~Mas hep-th/9805172 }

\lref\promise{N.~A.~Nekrasov and friends, to appear}

\lref\todalit{K.~Ueno, K.~Takasaki, Adv. Studies in Pure Math.
{\bf 4} (1984) 1 \semi For an excellent review see, e.g.
S.~Kharchev, hep-th/9810091}

 \Title{\vbox{\baselineskip 10pt \hbox{}
\hbox{ITEP-TH-22/02} \hbox{IHES/P/04/22} \hbox{hep-th/0206161}  }}
{\vbox{\vskip -30 true pt
\smallskip
   \centerline{SEIBERG-WITTEN PREPOTENTIAL}   \smallskip\smallskip\centerline{ FROM INSTANTON COUNTING}
\vskip4pt }} \vskip -20 true pt \centerline{Nikita
A.~Nekrasov\foot{on leave of absence from ITEP, 117259, Moscow,
Russia}}
\smallskip\smallskip
\smallskip\bigskip
 \centerline{\it
Institut des Hautes Etudes Scientifiques, Le Bois-Marie,
Bures-sur-Yvette, F-91440 France}
\medskip \centerline{\tt e-mail:
nikita@ihes.fr}
\bigskip

\ndt Direct evaluation of the Seiberg-Witten prepotential is
accomplished following the localization programme suggested in
\issues. Our results agree with all low-instanton calculations
available in the literature. We present a two-parameter
generalization of the Seiberg-Witten prepotential, which is rather
natural from the M-theory/five dimensional perspective, and
conjecture its relation to the tau-functions of KP/Toda hierarchy.
\bigskip
\bigskip
\centerline{\sl To Arkady Vainshtein on his 60th anniversary}

\Date{June 2002} \vfill\eject

\hfill{\vbox{\baselineskip12pt\sl \hbox{{\bf M.Jourdain}: \quad \
Par ma foi!} \hbox{\qquad\qquad\qquad\qquad il y a plus de
quarante ans}\hbox{\qquad\qquad\qquad\qquad que je dis de la
prose}\hbox{ \qquad\qquad\qquad\qquad sans que j'en suisse rien
;}\hbox{\qquad\qquad\qquad\qquad  et je vous suis le plus oblig\'e
du monde}\hbox{\qquad\qquad\qquad\qquad de m'avoir appris
cela....}}}

\bigskip\hfill{\vbox{\baselineskip 10pt\hbox{\it Le Bourgeois
gentilhomme.} \hbox{\sl J.~-B.~Moli\`ere}}}

\bigskip

\newsec{Introduction}
The dynamics of gauge theories is a long and fascinating subject.
The dynamics of supersymmetric gauge theories is a subject with
shorter history. However, more facts are known about susy theories, and with better precision \nsvz\ yet with rich
enough applications both in physics and mathematics. In
particular, the solution of Seiberg and Witten \sw\ of ${\CN}=2$
gauge theory using the constraints of special geometry of the
moduli space of vacua led to numerous achievements in
understanding of the strong coupling dynamics of gauge theory and
as well as string theory backgrounds of which the gauge theories
in question arise as low energy limits. The low energy effective
Wilsonian action for the massless vector multiplets $(a_l)$ is
governed by the prepotential $\CF (a; {\Lambda})$, which receives
one-loop perturbative and instanton non-perturbative corrections
(here $\Lambda$ is the dynamically generated scale):
\eqn\prepo{{\CF} (a; {\Lambda}) = {\CF}^{pert} ( a ; {\Lambda}) +
{\CF}^{inst} (a; {\Lambda})} In spite of the fact that these
instanton corrections were calculated in many indirect ways, their
gauge theory calculation is lacking beyond two
instantons\seibergone\twoinst. The problem is that the instanton
measure seems to get very complicated with the growth of the
instanton charge, and the integrals are hard to evaluate.

The present paper attempts to solve this problem via the
localization technique, proposed long time ago in
\issues\dbound\ihiggs. Although we tried to make the paper
readable to both mathematicians and physicists we don't expect it
to be quite understandable without some background material, which
we suggest to look up in \sw\tdgt\calculus.

\ndt {\bf Notations.} Let $G$ be a semi-simple Lie group, $T$ is
maximal torus, ${\bf g} = Lie(G)$ its Lie algebra, ${\bf t} =
Lie(T)$ its Cartan subalgebra, $W = N(T)/T$ denote its Weyl group,
${\CU} = ({\bf t} \otimes {\IC})/W$ denotes the complexified space
of conjugacy classes in ${\bf g}$. We consider the moduli space
$M_{k}(G)$ of framed $G$-instantons: the anti-self-dual gauge
fields $A$, $F_{A}^{+}=0$, in the principal $G$-bundle ${\CP}$
over the $4$-sphere ${\bf S}^4 = {\IR}^4 \cup {\infty}$ with
\eqn\pontr{k = -{1\over{8 h {\pi}^2}} \int_{{\IR}^4} \ {\tr} \
F_{A} \wedge F_{A}}considered up to the gauge transformations $g:
A \mapsto g^{-1} A g + g^{-1} dg$, s.t. $g({\infty}) = 1$. We also
consider several compactifications of the space $M_{k}(G)$: the
Uhlenbeck compactification ${\tilde M}_{k}(G)$ and the Gieseker
compactification ${\widetilde{\CM}_{k}}$ for $G = U(N)$ or
$SU(N)$. In the formula \pontr\ we use the trace in the adjoint
representation, and $h$ stands for the dual Coxeter number of $G$.

\item{} {\bf Field theory description.} We calculate vacuum expectation value of certain gauge
theory observables. These observables are annihilated by a
combination of the supercharges, and their expectation value is
not sensitive to various parameters, the energy scale in
particular. Hence, one can do the calculation in the ultraviolet,
where the theory is weakly coupled and the instantons dominate.
Or, one can do the calculation in the infrared, and relate the
answer to the prepotential of the effective low-energy theory. By
equating these two calculations we obtain the desired formula.

\item{} {\bf Mathematical description.}
We study $G \times {\bf T}^2$ equivariant cohomology of the moduli
space $\widetilde{\CM}_{k}$, where $G$ acts by rotating the gauge
orientation of the instantons at infinity, and ${\bf T}^2$ is the
maximal torus of $SO(4)$ -- the group of rotations of ${\IR}^4$
which also acts naturally on the moduli space\foot{Throughout the
paper we mostly consider the $SU(N)$ instantons (or $U(N)$
noncommutative instantons). We use the notation
$\widetilde{\CM}_{k,N}$ when we want to emphasize that the  gauge
group is $U(N)$.}. Let $p: {\widetilde{\CM}_{k}} \to pt$ be the
map collapsing the moduli space to a point. We consider the
following quantity: \eqn\local{Z (a , {\e}_1, {\e}_2 ; q )=
\sum_{k=0}^{\infty} q^k \oint_{\widetilde{\CM}_{k}} 1 } where
$\oint 1$ denotes the {\it localization} of the pushforward
$p_{*}1$ of $1 \in H^{*}_{G \times {\bf
T}^2}({\widetilde{\CM}_k})$ in $H^{*}_{G \times {\bf T}^2} ( pt )
= {\IC} [ {\CU}, {\e}_1, {\e}_2]$. We denote the coordinates on
$\bf t$ by $a$ and the coordinates on the Lie algebra of ${\bf
T}^2$ by ${\e}_1, {\e}_2$. In explicit calculations\foot{For $G=
SU(N)$ we actually use $a = (a_1, \ldots, a_N)$ s.t. $\sum_l a_l
=0$} we represent $1$ by a cohomologically equal form which allows
to replace $\oint 1$ by an ordinary integral:
\eqn\repr{\oint_{\widetilde{{\CM}_k}} 1 =
\int_{\widetilde{{\CM}_{k}}} {\exp} \ {\o} + {\m}_{G} (a) +
{\m}_{{\bf T}^2} ({\e}_1, {\e}_2) } where ${\o}$ is a symplectic
form on $\widetilde{{\CM}_{k}}$, invariant under the $G \times
{\bf T}^2$ action, and ${\m}_{G}, {\m}_{{\bf T}^2}$ are the
corresponding moment maps.

\ndt  Our first claim is \eqn\prep{\mathboxit{Z ( a, {\e}_1,
{\e}_2 ; q) = {\exp} \left( {{\CF}^{inst} (a, {\e}_1, {\e}_2; q)
\over {\e}_1 {\e}_2} \right)}} where the function ${\CF}^{inst}$
is analytic in ${\e}_1, {\e}_2$ near ${\e}_1 = {\e}_2 = 0$.

\ndt We also have the following explicit expression for $Z$ in the
case\foot{in the general case we also have a formula, but it looks
less transparent}  ${\e}_1 = - {\e}_2 = {\hbar}$ for\foot{a simple
generalization to $SO$ and $Sp$ cases will be presented in
\promise} $G = SU(N)$ : \eqn\explct{\mathboxit{Z ( a, {\hbar},
-{\hbar}; q) = \sum_{{\vec {\bf k}}} q^{\vert {\bf k} \vert}
\prod_{(l,i) \neq (n,j) } {{a_{ln} + {\hbar} \left( k_{l,i} -
k_{n,j} +j - i \right)}\over{a_{ln} + {\hbar} \left( j - i
\right)}}}}Here $a_{ln} = a_l - a_n$, the sum is over all colored
partitions: $\vec{\bf k} = \left( {\bf k}_1, \ldots, {\bf k}_N
\right)$, ${\bf k}_{l} = \{ k_{l,1} \geq k_{l,2} \geq \ldots
k_{l,n_l} \geq k_{l,n_{l}+1} = k_{l, n_{l}+2} = \ldots = 0 \}$,
$$\vert {\vec {\bf k}} \vert = \sum_{l, i} k_{l,i}  \ ,$$ and the
product is over $1\leq l,n \leq N$, and $ i,j \geq 1$.

\ndt Already \explct\ can be used to make rather powerful checks
of the Seiberg-Witten solution. But the checks are more impressive
when one considers the theory with fundamental matter. To get
there one studies the bundle $V$ over ${\widetilde{{\CM}_k}}$ of
the solutions of the Dirac equation in the instanton background.
Let us consider the theory with $N_f$ flavors. It can be shown
that the gauge theory instanton measure calculates in this case
(cf. \instmeasures): \eqn\instmat{Z (a, m, {\e}_1, {\e}_2; q) =
\sum_{k} q^{k} \oint_{\widetilde{{\CM}_k}} {\rm Eu}_{G \times {\bf
T}^2\times U(N_f)} ( V \otimes M )} where $M = {\IC}^{N_f}$ is the
flavor space, where acts the flavor group $U(N_f)$, $m = (m_1,
\ldots, m_{N_f})$ are the masses = the coordinates on the Cartan
subalgebra of the flavor group Lie algebra, and finally ${\rm
Eu}_{G \times {\bf T}^2 \times U(N_f)}$ denotes the equivariant
Euler class.

\ndt The formula \explct\ generalizes in this case to:
\eqn\explctm{\eqalign{& Z (a, m, {\e}_1, {\e}_2; q) = \sum_{\vec
{\bf k}} \left( q {\hbar}^{N_f} \right)^{\vert {\bf k} \vert}
\prod_{(l,i)} \prod_{f=1}^{N_f} {{{\Gamma} ( {a_l + m_f \over
{\hbar}} + 1 + k_{l,i} - i )}\over{{\Gamma} ( {a_l + m_f \over
\hbar} + 1 - i )}} \  \times \cr & \qquad \qquad \qquad \qquad
\qquad \qquad \qquad \qquad \qquad \times \ \prod_{(l,i) \neq
(n,j) } {{a_{ln} + {\hbar} \left( k_{l,i} - k_{n,j} +j - i
\right)}\over{a_{ln} + {\hbar} \left( j - i \right)}}\cr}} Again,
we claim that \eqn\analytic{{\CF}^{inst}( a, m, {\e}_1, {\e}_2; q)
= {\e}_1 {\e}_2 \ {\rm log} Z ( a, m, {\e}_1, {\e}_2; q)} is
analytic in ${\e}_{1,2}$.

\ndt The formulae \explct\explctm\ were checked against the
Seiberg-Witten solution \swsol. Namely, we claim that \bigskip
\boxit{\bigskip ${\CF}^{inst} ( a, m , {\e}_1, {\e}_2)
\vert_{{\e}_1 = {\e}_2 = 0} = $ the instanton part of the
prepotential of the low-energy effective theory of the ${\CN}=2$
gauge theory with the gauge group $G$ and $N_f$ fundamental matter
hypermultiplets.}

\item{} {\bf Mathematical formulation.} The latter statement
means that ${\CF}^{inst}$ is related to periods of a family of
curves. More precisely, consider the following family of
curves\foot{${\Lambda}, m_f$ are fixed for the family}
${\Sigma}_{u}$ (here we formulate things for $G = SU(N)$ but the
generalization to general $G$ is well-known \swsol\ ): \eqn\crvs{
w + {{\Lambda}^{2N - N_f} Q ( {\l} ) \over w} = {\bf P} ( {\l} ) =
\prod_{l=1}^{N} ({\l}- {\a}_l)}where $Q ({\l}) = \prod_{f=1}^{N_f}
( {\l} + m_f )$. The base of the family \crvs\ is the space ${\CU}
= {\IC}^{N-1} \ni u$ of the polynomials ${\bf P}$ (we set $\sum_l
{\a}_l = 0$). Consider the region ${\CU}^{pert} \subset {\CU}$
where $\vert {\a}_l \vert, \vert {\a}_l - {\a}_n \vert \gg \vert
{\Lambda}\vert, \vert m_f\vert$. In ${\CU}^{pert}$ we can pass
from the local coordinates $( {\a}_l )$ to the local coordinates
$( a_l )$ given by: \eqn\prd{a_l = {1\over{2\pi i}} \oint_{A_l} \
{\l} {{dw} \over w}}where the cycle $A_l$ can be described as
encircling the cut on the ${\l}$ plane connecting the points
${\a}_l^{\pm} = {\a}_l + o ({\Lambda}) $ which solve the
equations: \eqn\cuts{\pm 2 {\Lambda}^{N- {{N_f}\over 2}}
Q^{\half}({\a}_l^{\pm}) = {\bf P} ({\a}_l^{\pm})} The sum $\sum_l
{A_l}$ vanishes in the homology of $\Sigma_{u}$, therefore we get
$N-1$ independent coordinates, as we should have. Now, define the
dual coordinates \eqn\dualc{a_l^{D} = {1\over{2\pi i}} \oint_{B_l}
\ {\l} {{dw} \over w}} where $B_l$ encircles the cut connecting
${\a}_l^{+}$ and ${\a}_{(l+1) mod N }^{-}$. Then, one can show
\sw\swsol\ that
$$
\sum_l d a_l \wedge d a_l^D = 0
$$ on ${\CU}$, and as a consequence, there exists a
(locally defined) function, called prepotential, ${\CF} (a; m,
{\Lambda})$ such that \eqn\prepo{\sum_l \  a_l^D {\rm d} a_l =
{\rm d} {\CF} ( a)} In the region ${\CU}^{pert}$ the prepotential
has the expansion: \eqn\expns{\eqalign{& {\CF} (a; m , \Lambda) =
{\CF}^{pert} (a) + {\CF}^{inst}(a) \cr &  {\CF}^{pert} (a) =
{\half} \sum_{l \neq n} \left( a_l - a_n \right)^2 {\rm log}
\left( {a_l - a_n \over {\Lambda}} \right) - \sum_{l,f} ( a_l +
m_f )^2 {\rm log} \left( {{a_l + m_f}\over{\Lambda}} \right)\cr}}
where ${\CF}^{inst}$ is a power series in ${\Lambda}$. Our claim
is that
\bigskip\boxit{\bigskip ${\CF}^{inst}$ defined by the formula \expns\ coincides
with ${\CF}^{inst} ( a, m , {\e}_1, {\e}_2 ) \vert_{{\e}_1 =
{\e}_2 = 0 }$.}
\bigskip

\ndt We have checked this claim by an explicit calculation for up
to five instantons, against the formulae in \experiment.

\ndt There is also a generalization of \explct\ to the case of
adjoint matter. It is presented in the main body of the paper.

\centerline{* \qquad * \qquad *}

\ndt This paper is a short version of a longer manuscript
\promise, which will contain various details. In this paper we
mostly state the results.

\ndt {\bf The paper is organized as follows.} In the next section
we describe the physical idea of our calculation. We define the
observable of interest, and sketch two calculations of its
expectation value -- in the weak coupling regime in the
ultraviolet, and the infrared low-energy effective theory
calculation. The section $3$ provides more details on the
instanton calculation and generalizes the pure gauge theory
calculation to the case of the theories with matter. We also
discuss explicit low instanton charge calculations. In the section
$4$ we discuss our results from the M-theory viewpoint, consider
some generalizations, present our conjectures and describe future
directions.

\ndt{\bf Acknowledgements.} This paper would have never seen the
light without the numerous conversations of the author with
A.~Losev. We also benefited from discussions/collaborations with
A.~Givental, G.~Moore, A.~Okounkov, S.~Shatashvili, A.~Vainshtein,
H.~Braden, S.~Cherkis, K.~Froyshov, V.~Kazakov, I.~Kostov,
A.~Marshakov and A.~Morozov over the last five years. We are
especially grateful to T.~Hollowood for reading the manuscript and
sending us his comments, in particular for pointing out an
important typo.

We are most grateful to T.~Piatina for providing the opportunity
to accomplish this work, and for inspiring us during the difficult
moments of research (especially between the third and the fourth
instantons).

Research was supported in part by {\cyr RFFI} grant 01-01-00549
and by the grant 00-15-96557 for the support of scientific
schools.

The results of this paper were presented at the EURESCO school
``Particle physics and gravitation'' held at Bad Herrenalb. We
thank H.~Nicolai for the invitation and for organizing a nice
school.

\newsec{Field theory expectations}

In this section we explain our approach in the field theory
language. We exploit the fact that the supersymmetric gauge theory
on flat space has a large collection of observables whose
correlation functions are saturated by instanton contribution in
the limit of weak coupling. In addition, in the presence of the
adjoint scalar vev these instantons tend to shrink to zero size.
Moreover, the observables we choose have the property that the
instantons which contribute to their expectation values are
localized in space. This solves the problem of the runaway of
point-like instantons, pointed out in \issues.

\subsec{Supersymmetries and twisted supersymmetries}

The ${\CN}=2$ theory has eight conserved supercharges, $Q_{\a}^i,
Q_{\dot a}^i$, which transform under the global symmetry group
$SU(2)_L \times SU(2)_R \times SU(2)_I$ of which the first two
factors belong to the group of spatial rotations and the last one
is the $R$-symmetry group. The indices ${\a}, {\dot\a}, i$ are the
doublets of these respective $SU(2)$ factors. The basic multiplet
of the gauge theory is the vector multiplet. Here is the spin
content of its members: $$ \eqalign{ {\bf Field} & {\quad SU(2)_L
\quad SU(2)_R \quad SU(2)_I} \cr \underline{\qquad\quad} &
\underline{\qquad\qquad\qquad\qquad\qquad\qquad\qquad} \cr A_{\m}
& \qquad \half \qquad\qquad \half \qquad\qquad 0 \cr \psi^{i}_{\a}
& \qquad \half \qquad\qquad 0 \qquad\qquad \half \cr
\psi^{i}_{\dot\a} & \qquad 0 \qquad\qquad \half \qquad\qquad \half
\cr \phi, \bar\phi & \qquad 0 \qquad\qquad 0 \qquad\qquad 0 \cr}
$$ It is useful to work in the notations which make only $SU(2)_L
\times SU(2)_d$ part of the global symmetry group manifest. Here
$SU(2)_d$ is the diagonal subgroup of $SU(2)_R \times SU(2)_I$. If
we call this subgroup a ``Lorentz group'', then the supercharges,
superspace, and the fermionic fields of the theory split as
follows:

\ndt{\bf Fermions}: ${\psi}_{\m}, {\chi}^{+}_{\m\n}, \eta$\semi
\ndt{\bf Superspace}: ${\t}^{\m}, {\bar \t}_{\m\n}^{+}, {\bar
\t}$\semi \ndt{\bf Superfield:} $ {\Phi} = {\phi} + {\t}^{\m}
{\psi}_{\m} + {\half} {\t}^{\m} {\t}^{\n} F_{\m\n} + \ldots $\semi
\ndt{\bf Supercharges}: $Q, Q_{\m\n}^{+}, G_{\m}$.

\ndt The supercharge $Q$ is a scalar with respect to the ``Lorentz
group'' and is usually considered as a BRST charge in the
topological quantum field theory version of the susy gauge theory.
It is conserved on any four-manifold.

In \wittenone\ E.~Witten has employed a self-dual two-form
supercharge $Q_{\m\n}^{+}$ which is conserved on K\"ahler
manifolds.

Our idea is to use other supercharges $G_{\m}$ as well. Their
conservation is tied up with the isometries of the four-manifold
on which one studies the gauge theory. Of course, the idea to
regularize the supersymmetric theory by subjecting it to the
twisted boundary conditions is very common both in physics
\twists, and in mathematics
\torusaction\maxim\givental\torusaction.

At this point we should mention that the idea to apply
localization techniques to the instanton integrals has been
recently applied in \hollowood\ in the one- and two-instanton
cases. Without ${\bf T}^2$-localization this is still rather
complicated, yet simpler, calculation then the direct evaluation
\twoinst. We refer the interested reader to the beautiful review
\calculus\ for more details.

\subsec{Good observables: UV}

In the applications of the susy gauge theory to Donaldson theory,
where one works with the standard topological supercharge $Q$, the
observables one is usually interested in are the gauge invariant
polynomials ${\CO}^{(0)}_{P,x} = P({\phi}(x))$ in the adjoint
scalar ${\phi}$, evaluated at space-time point $x$, and its
descendants: ${\CO}^{(k)}_{P, {\Sigma}} = \int_{\Sigma} P (
{\Phi})$, where $\Sigma$ is a $k$-cycle. Unfortunately for $k
> 0$ all such cycles are homologically trivial on ${\IR}^4$ and no
non-trivial observables are constructed in such a way. One
construct an equivalent set of dual observables by integration
over ${\IR}^4$ of a product of a closed $k$-form  ${\o} =
{1\over{(4-k)!}} \ {\o}_{{\m}_1 \ldots {\m}_{k}} {\t}^{{\m}_1}
\ldots {\t}^{{\m}_{k}} $ and the $4-k$-form part of $P({\Phi})$:
\eqn\dualo{{\CO}_{P}^{\omega} = \int d^4 x d^4 {\t} \ {\o} (x,
{\t}) \  P ( {\Phi} )} Again, most of these observables are
$Q$-exact, as any closed $k$-form on ${\IR}^4$ is exact for $k
> 0$.

However, if we employ the rotational symmetries of ${\IR}^4$ and
work equivariantly, we find new observables.

\ndt Namely, consider the fermionic charge \eqn\rotch{{\tilde Q} =
Q +
 E_{a} {\Omega}^{a}_{\m\n} x^{\n} G_{\m}} Here ${\Omega}^a =
{\Omega}^{a}_{\m\n} x^{\n} {\p}_{\m}$ for $a=1 \ldots 6$ are the
vector fields generating $SO(4)$ rotations, and $E \in Lie
(SO(4))$ is a formal parameter.

\ndt With respect to the charge ${\tilde Q}$ the observables
${\CO}^{(k)}_{P,\Sigma}$ are no longer invariant\foot{except for
${\CO}^{(0)}_{P, 0}$ where $0 \in {\IR}^4$ is the origin, left
fixed by the rotations}. However, the observables \dualo\ can be
generalized to the new setup, producing a priori nontrivial
${\tilde Q}$-cohomology classes. Namely, let us take any
$SO(4)$-equivariant form on ${\IR}^4$. That is, take an
inhomogeneous differential form ${\Omega} (E)$ on ${\IR}^4$ which
depends also on an auxiliary variable $E \in Lie (SO(4))$ which
has the property that for any $g \in SO(4)$: \eqn\equiv{g^*
{\Omega} (E) = {\Omega} ( g^{-1} E g )}where we take pullback
defined with the help of the action of $SO(4)$ on ${\IR}^4$ by
rotations. Such $E$-dependent forms are called equivariant forms.
On the space of equivariant forms acts the so-called equivariant
differential, \eqn\equivd{D = d + \iota_{V(E)}} where $V(E)$ is
the vector field on ${\IR}^4$ representing the infinitesimal
rotation generated by $E$. For equivariantly closed (i.e.
$D$-closed) form ${\Omega}(E)$ the observable:
\eqn\equivob{{\CO}_{P}^{{\Omega}(E)} = \int_{{\IR}^4} {\Omega} (E)
\wedge P ( {\Phi}) } is ${\tilde Q}$-closed.

\ndt Any $SO(4)$ invariant polynomial in $E$ is of course an
example of the $D$-closed equivariant form. Such a polynomial is
characterized by its restriction onto the Cartan subalgebra of
$SO(4)$, where it must be Weyl-invariant. The Cartan subalgebra of
$SO(4)$ is two-dimensional. Let us denote the basis in this
subalgebra corresponding to the decomposition ${\IR}^4 = {\IR}^2
\oplus {\IR}^2$ into a orthogonal direct sum of two dimensional
planes, by $({\e}_1, {\e}_2)$. Under the identification $Lie
(SO(4)) \approx Lie (SU(2)) \oplus Lie (SU(2))$ these map to
$({\e}_1 + {\e}_2, {\e}_1 - {\e}_2)$. The Weyl (= ${\IZ}_2 \times
{\IZ}_2$) invariant polynomials are polynomials in ${\s} =
{\e}_1^2 + {\e}_2^2$ and ${\chi} = {\e}_1 {\e}_2$. As $SO(4)$ does
not preserve any forms except for constants we should relax the
$SO(4)$ symmetry to get interesting observables.

Thus, let us fix in addition a translationally invariant
symplectic form ${\o}$ on ${\IR}^4$. Its choice breaks $SO(4)$
down to $U(2)$ -- the holonomy group of a K\"ahler manifold. Let
us fix this $U(2)$ subgroup. Then we have a moment map:
\eqn\mmnt{{\m}: {\IR}^4 \longrightarrow Lie(U(2))^*, \qquad d
{\m}(E) = \iota_{V(E)} {\o}, \quad E \in Lie(U(2))} And therefore,
the form ${\o} - {\m} (E)$ is $D$-closed. One can find such
euclidean coordinates $x^{\n}$, ${\n} = 1, 2,3,4$ that the form
${\o}$ reads as follows: \eqn\sympl{{\o} = dx^1 \wedge dx^2 + dx^3
\wedge dx^4} The Lie algebra of $U(2)$ splits as a direct sum of
one-dimensional abelian Lie algebra of $U(1)$ and the Lie algebra
of $SU(2)$. Accordingly, the moment map ${\m}$ splits as $( h,
{\m}^1, {\m}^2, {\m}^3)$. In the $x^\m$ coordinates \eqn\mmntc{h =
\sum_{\m} \left( x^{\m} \right)^2 , \qquad {\m}^a = {\half}
{\eta}^{a}_{\m\n} x^{\m} x^{\n} ,} where ${\eta}^a_{\m\n}$ is the
anti-self-dual 't Hooft symbol.

\ndt Finally, the choice of ${\o}$ also defines a complex
structure on ${\IR}^4$, thus identifying it with ${\IC}^2$ with
complex coordinates $z_1, z_2$ given by: $z_1 = x^1 + i x^2$, $z_2
= x^3 + i x^4$. For $E$ in the Cartan subalgebra $H = {\m}(E)$ is
given by the simple formula: \eqn\mcart{H = {\e}_1 \vert z_1
\vert^2 + {\e}_2 \vert z_2 \vert^2} After all these preparations
we can define the correlation function of our interest:
\eqn\mastercorr{Z ( a, {\e}) = \left\langle {\exp} {1\over{(2\pi
i)^2}} \int_{{\IR}^4} \left( {\o} \wedge {\Tr} \left( {\phi} F +
{\half} {\psi} {\psi} \right) - H \ {\Tr} \left( F \wedge F
\right) \right)\right\rangle_{a} } where we have indicated that
the vacuum expectation value is calculated in the vacuum with the
expectation value of the scalar ${\phi}$ in the vector multiplet
given by $a \in {\bf t}$. More precisely, $a$ will be the central
charge of ${\CN}=2$ algebra corresponding to the $W$-boson states
(cf.\sw) .

\ndt{\bf Remarks.} \item{} 1.) Note that the observable in
\mastercorr\ makes the widely separated instantons suppressed.
More precisely, if the instantons form clusters around points
${\vec r}_1, \ldots, {\vec r}_l$ then they contribute $\sim {\exp}
- \sum_{m} H ({\vec r}_m )$ to the correlation function.

\item{} 2.) One can expand \mastercorr\ as a sum over different instanton
sectors: $$ Z ( a, {\e} ) = \sum_{k=0}^{\infty} q^{k} Z_{k} (a,
{\e})$$ where $q \sim {\Lambda}^{2N}$ is the dynamically generated
scale -- for us -- simply the generating parameter.

\item{} 3.) The observable \mastercorr\ is formally cohomologous
to the identity, as $$\int \left( {\o} + H \right) \ {\Tr}
{\Phi}^2 = {\tilde Q} \int d^4 x d^4 {\t} \ {\bf A} ( x, {\t} )  \
{\Tr} {\Phi}^2 \ , $$where ${\bf A}(x,{\t}) = {\o}_{\m\n} x^{\m}
{\t}^{\n}$. We cannot eliminate it without having to perform the
full path integral, however, as it serves as a supersymmetric
regulator. On the other hand,  in the presence of this observable
the path integral can be drastically simplified, the fact we shall
exploit below. The analogous manipulation in the context of two
dimensional supersymmetric Landau-Ginzburg models was done in the
first reference in  \twists.

\ndt The supersymmetry guarantees that \mastercorr\ is saturated
by instantons. Moreover, the superspace of instanton zero modes is
acted on by a finite dimensional version of the supercharge
${\tilde Q}$ which becomes an equivariant differential on the
moduli space of framed instantons. Localization with respect to
this supercharge reduces the computation to the counting of the
isolated fixed points and the weights of the action of the
symmetry groups (a copy of gauge group and $U(2)$ of rotations) on
the tangent spaces. This localization can be understood as a
particular case of the Duistermaat-Heckman formula \DHf, as
\mastercorr\ calculates essentially the integral of the exponent
of the Hamiltonian of a torus action (Cartan of $G$ times ${\bf
T}^2$) against the symplectic measure.

\ndt The counting of fixed points can be nicely summarized by a
contour integral (see below). This contour integral also can be
obtained by transforming the integral over the ADHM moduli space
of the observable \mastercorr\ evaluated on the instanton
configuration, by adding ${\tilde Q}$-exact terms, as in
\ihiggs\dbound. It also can be derived from Bott's formula \bott.

\subsec{Good observables: IR}

The nice feature of the correlator \mastercorr\ is it simple
relation to the prepotential of the low-energy effective theory.
In order to derive it let us think of the observable \mastercorr\
as of a slow varying changing of the microscopic coupling
constant. If we could completely neglect the fact that $H$ is not
constant, then its addition would simply renormalize the effective
low-energy scale ${\Lambda} \to {\Lambda} e^{-H}$.

However, we should remember that $H$ is not constant, and regard
this renormalization as valid up to terms in the effective action
containing derivatives of $H$. Moreover, $H$ is really a bosonic
part of the function ${\CH} (x, {\t})$ on the (chiral part of)
superspace (in \issues\ such superspace-dependent deformations of
the theory on curved four-manifolds were considered): $$ {\CH} (
x, {\t}) = H (x) + {\half} {\o}_{\m\n} {\t}^{\m} {\t}^{\n} $$
Together these terms add up to the making the standard
Seiberg-Witten effective action determined by the prepotential
${\CF} (a; {\Lambda})$ to the one with the superspace-dependent
prepotential \eqn\swpre{{\CF} (a; {\Lambda} e^{-{\CH}(x, {\t})}) =
{\CF} ( a ; {\Lambda} e^{-H} ) + {\o} \ {\Lambda}{\p}_{\Lambda}
{\CF} (a; {\Lambda} e^{-H}) + {\half} {\o}^2 \ \left( {\Lambda}
{\p}_{ \Lambda} \right)^2 {\CF} (a; {\Lambda} e^{-H})} This
prepotential is then integrated over the superspace (together with
the conjugate terms) to produce the effective action.

Now, let us go to the extreme infrared, that is let us scale the
metric on ${\IR}^4$ by a very large factor $t$ (keeping ${\o}$
intact). On flat ${\IR}^4$ the only term which may contribute to
the correlation function in question in the limit $t \to \infty$
is the last term in \mastercorr\ as the rest will (after
integration over superspace) necessarily contain couplings to the
gauge fields which will require some loop diagrams to get
non-trivial contractions, which all will be suppressed by inverse
powers of $t$. The last term, on the other hand, gives:

\eqn\masterir{Z(a; {\e}) = {\exp} - {1\over
8{\pi}^2}\int_{{\IR}^4} {\o} \wedge {\o} {{{\p}^2 \ {\CF} (a;
{\Lambda} e^{-H})}\over{( {\p} {\rm log} {\Lambda})^2}}  + O
({\e})} where we used the fact that the derivatives of $H$ are
proportional to ${\e}_{1,2}$. Recalling \sympl\mmntc\ the integral
in \masterir\ reduces to: \eqn\mastercorrir{Z ( a; {\e}_1, {\e}_2
) = {\exp}\ {{\CF}^{inst}(a ; {\Lambda}) + O ({\e}) \over {{\e}_1
{\e}_2}}} where $${\CF}^{inst} (a; {\Lambda})  = \int_{0}^{\infty}
{\p}_{H}^2 {\CF} (a; {\Lambda} e^{-H}) \ H {\rm d} H $$ thereby
explaining our claim about the analytic properties of $Z$ and
${\CF}$.

\newsec{Instanton measure and its localization}

\subsec{ADHM data}

The moduli space ${\CM}_{k,N}$ of instantons with fixed framing at
infinity has dimension $4kN$. It has the following convenient
description. Take two complex vector spaces $V$ and $W$ of the
complex dimensions $k$ and $N$ respectively. These spaces should
be viewed as Chan-Paton spaces for $D(p-4)$ and $Dp$ branes in the
brane realization of the gauge theory with instantons.

Let us also denote by $L$ the two dimensional complex vector
space, which we shall identify with the Euclidean space ${\IR}^4
\approx {\IC}^2$ where our gauge theory lives.

Then the ADHM \adhm\ data consists of the following maps between
the vector spaces: \eqn\sqnc{V \longrightarrow^{\kern -.15in
{\tau}} \quad V {\otimes} L \oplus W \longrightarrow^{\kern -.15in
{\s}} \quad V \otimes {\Lambda}^2 L} where \eqn\tausigma{\eqalign{
& {\tau} = \pmatrix{ B_2 \cr - B_1 \cr J}, \qquad {\s} = \pmatrix{
B_1 & B_2 & I} \cr & \cr & B_{1,2} \in {\rm End} (V), \ I \in {\rm
Hom} ( W, V ), \ J \in {\rm Hom} (V, W) \cr}}

The ADHM construction represents the moduli space of $U(N)$
instantons on ${\IR}^4$ of charge $k$ as a hyperk\"ahler quotient
\rkh\  of the space of operators $(B_1, B_2, I, J)$ by the action
of the group $U(k)$ for which $V$ is a fundamental representation,
$B_{1},B_{2}$ transform in the adjoint, $I$ in the fundamental,
and $J$ in the anti-fundamental representations.

More precisely, the moduli space of proper instantons is obtained
by taking the quadruples $(B_{1,2}, I, J)$ obeying the so-called
ADHM equations: \eqn\mmntmps{{\m}_c = 0, \qquad {\m}_r = 0
,}where: \eqn\adhmeq{\eqalign{& {\m}_c = [B_1, B_2] + IJ \cr
{\m}_r = & [B_1, B_1^{\dagger}] + [ B_2, B_2^{\dagger}] +
II^{\dagger} - J^{\dagger} J \cr}} and with the additional
requirement that the stabilizer of the quadruple in $U(k)$ is
trivial. This produces a non-compact hyperk\"ahler manifold
$M_{k,N}$ of instantons with fixed framing at infinity.

\ndt The framing is really the choice of the basis in $W$. The
group $U(W) = U(N)$ acts on these choices, and acts on $M_{k,N}$,
by transforming $I$ and $J$ in the anti-fundamental and the
fundamental representations respectively.

\ndt This action also preserves the hyperk\"ahler structure of
$M_{k,N}$ and is generated by the hyperk\"ahler moment maps:
\eqn\glbmmps{{\bf m}_r = I^{\dagger} I - J J^{\dagger}, \qquad
{\bf m}_c = J I} Actually, ${\Tr}_{W} {\bf m}_{r,c} = {\Tr}_{V}
{\m}_{r,c}$, thus the central $U(1)$ subgroup of $U(N)$ acts
trivially on $M_{k,N}$. Therefore it is the group $G =
SU(N)/{\IZ}_N$ which acts non-trivially on the moduli space of
instantons.

\subsec{Instanton measure}

The supersymmetric gauge theory measure can be regarded as an
infinite-dimensional version of the equivariant Matthai-Quillen
representative of the Thom class of the bundle ${\Gamma} \left(
{\Omega}^{2,+} \otimes {\bf g}_P\right)$ over the
infinite-dimensional space of all gauge fields ${\CA}_{P}$ in the
principal $G$-bundle $P$ (summed over the topological types of
$P$). In physical terms, in the weak coupling limit we are
calculating the supersymmetric delta-function supported on the
instanton gauge field configurations. In the background of the
adjoint Higgs vev, this supersymmetric delta-function is actually
an equivariant differential form on the moduli space $M_{k,N}$ of
instantons. It can be also represented using the
finite-dimensional hyperk\"ahler quotient ADHM construction of
$M_{k,N}$ (as opposed to the infinite-dimensional quotient of the
space of all gauge fields by the action of the group of gauge
transformations, trivial at infinity) \ihiggs: \eqn\insm{ \int
{\CD} {\phi} {\CD} {\bar\phi} {\CD} {\vec H} {\CD} {\vec\chi}
{\CD} {\eta} {\CD} {\Psi} {\CD} B {\CD} I {\CD} J \ e^{{\tilde Q}
\left( {\vec\chi} \cdot {\vec \m} (B, I, J) + {\Psi} \cdot V
\left( {\bar\phi} \right) + {\eta} [ {\phi}, {\bar\phi} ]
\right)}}where, say: \eqn\qtild{\eqalign{& {\tilde Q} B_{1,2} =
{\Psi}_{B_{1,2}}, \quad {\tilde Q} {\Psi}_{B_{1,2}} =  [ {\phi} ,
B_{1,2}] + {\e}_{1,2} B_{1,2} \cr & {\tilde Q} I = {\Psi}_{I},
\quad {\tilde Q} {\Psi}_{I} = {\phi} I - I a \cr & {\tilde Q} J =
{\Psi}_{J}, \quad {\tilde Q} {\Psi}_{J} = - J {\phi} + J a -
({\e}_1 + {\e}_2) J \cr & {\tilde Q} {\chi}_r = H_{r}, \ {\tilde
Q} H_r = [ {\phi}, {\chi}_r ], \qquad {\tilde Q} {\chi}_c = H_c, \
{\tilde Q} H_c = [ {\phi}, \chi_{c} ] + ( {\e}_1 + {\e}_2 )
{\chi}_c \cr & {\Psi} \cdot V \left( {\bar\phi} \right) = {\Tr}
\left( {\Psi}_{B_{1}} [ {\bar\phi}, B_{1}^{\dagger}] +
{\Psi}_{B_{2}} [ {\bar\phi}, B_{2}^{\dagger}] + {\Psi}_{I}
[{\bar\phi}, I^{\dagger}] - {\Psi}_{J} [ {\bar \phi}, J^{\dagger}]
+ c.c. \right) \cr}} (we refer to \ihiggs\ for more detailed
explanations). If the moduli space $M_{k,N}$ was compact and
smooth one could interpret \insm\ as a certain topological
quantity and apply the powerful equivariant localization
techniques \tdgt\ to calculate it.

\ndt The non-compactness of the moduli space of instantons is of
both ultraviolet and of infrared nature. The UV non-compactness
has to do with the instanton size, which can be made arbitrarily
small. The IR non-compactness has to do with the non-compactness
of ${\IR}^4$ which permits the instantons to run away to infinity.

\subsec{Curing non-compactness}

The UV problem can be solved by relaxing the condition on the
stabilizer, thus adding the so-called point-like instantons. A
point of the hyperk\"ahler space ${\tilde M}_{k,N}$ with orbifold
singularities which one obtains in this way (Uhlenbeck
compactification) is an instanton of charge $p \leq k$ and a set
of $k-p$ points on ${\IR}^4$: \eqn\uhl{{\tilde M}_{k,N} = M_{k,N}
\cup M_{k-1,N} \times {\IR}^4 \cup M_{k-2,N} \times Sym^2
({\IR}^4) \cup \ldots \cup Sym^k ({\IR}^4)}

The resulting space ${\tilde M}_{k,N}$ is a geodesically complete
hyperk\"ahler orbifold. Its drawback is the non-existence of the
universal bundle with the universal instanton connection over
${\tilde M}_{k,N} \times {\IR}^4$. We actually think that in
principle one can still work with this space. Fortunately, in the
case of $U(N)$ gauge group there exists a nicer space
$\widetilde{\CM}_{k,N}$ which is obtained from ${\tilde M}_{k,N}$
by a sequence of blowups (resolution  of singularities) which is
smooth, and after some modification of the gauge theory
(noncommutative\connes\cds\witsei\ deformation) becomes a moduli
space with the universal instanton. Technically this space is
obtained \neksch\ by the same ADHM construction except that now
one performs the hyperk\"ahler quotient at the non-zero level of
the moment map: \eqn\dfm{ {\m}_r = {\z}_r {\bf 1}_{V}, \qquad
{\m}_c = 0 } (one can also make ${\m}_c \neq 0$ but this does not
give anything new). The space of quadruples $(B_1, B_2, I,J)$
obeying \dfm\ is freely acted on by $U(k)$. The cohomology theory
of $\widetilde{\CM}_{k,N}$ is richer then that of ${\tilde
M}_{k,N}$ because of the exceptional divisors. However, our goal
is to study the original gauge theory. Therefore we are going to
consider the (equivariant) cohomology classes of
$\widetilde{\CM}_{k,N}$ lifted from ${\tilde M}_{k,N}$.

As we stated in the introduction, we are going to utilize the
equivariant symplectic volumes of $\widetilde{\CM}_{k,N}$. This is
not quite precise. We are going to consider the symplectic
volumes, calculated using the closed two-form lifted from ${\tilde
M}_{k,N}$. This form vanishes when restricted onto the exceptional
variety. This property ensures that we don't pick up anything not
borne in the original gauge theory (don't pick up freckle
contribution in the terminology of \freck).

The ADHM construction from the previous section gives rise to the
instantons with fixed gauge orientation at infinity (fixed
framing). The group $G = SU(N)/{\IZ}_N$ acts on their moduli space
${\CM}_{N,k}$ by rotating the gauge orientation. Also, the group
of Euclidean rotations of ${\IR}^4$ acts on ${\CM}_{N,k}$. We are
going to apply localization techniques with respect to both of
these groups.

In fact, it is easier to localize first with respect to the groups
$U(k) \times G \times {\bf T}^2$ acting on the vector space of
ADHM matrices, and then integrate out the $U(k)$ part of the
localization multiplet, to incorporate the quotient.

The action of ${\bf T}^2$ is free at ``infinities'' of
${\widetilde {\CM}_{k}}$, thus allowing to apply localization
techniques without worrying about the IR non-compactness.
Physically, the integral \mastercorr\ is Gaussian-like and
convergent in the IR region.

\subsec{Reduction to contour integrals}

After the manipulations as in \ihiggs\dbound\ we end up with the
following integral\freck: \eqn\cntr{Z_{k}( a, {\e}_{1}, {\e}_2)  =
{1\over{k!}} {{\e}^k\over{(2{\pi} i {\e}_1 {\e}_2)^k}} \oint
\prod_{I=1}^{k} {{\rm d}{\phi}_I \ Q({\phi}_I) \over {P({\phi}_I)
P({\phi}_I + {\e})}} \ {\prod}_{1\leq I < J \leq k}
{{{\phi}_{IJ}^2 ( {\phi}_{IJ}^2 - {\e}^2)}\over{({\phi}_{IJ}^2 -
{\e}_1^2)({\phi}_{IJ}^2- {\e}_2^2)}}} where: \eqn\poly{\eqalign{&
Q(x) = \prod_{f=1}^{N_f} ( x + m_f) \cr & P (x) = \prod_{l=1}^{N}
( x - a_l), \cr}}${\phi}_{IJ}$ denotes ${\phi}_I - {\phi}_J$, and
${\e} = {\e}_1 + {\e}_2$.

We went slightly ahead of time and presented the formula which
covers the case of the gauge theory with $N_f$ fundamental
multiplets. In fact, its derivation is rather simple if one keeps
in mind the relation to the Euler class of the Dirac zeromodes
bundle over the moduli space of instantons, stated in the
introduction.

The integrals \cntr\ should be viewed as contour integrals. As
explained in \dbound\ the poles at ${\phi}_{IJ} = {\e}_1, {\e}_2$
should be avoided by shifting ${\e}_{1,2} \to {\e}_{1,2} + i0$,
those at ${\phi}_I = a_l$ similarly by $a_l \to a_l + i0$ (this
case was not considered in \dbound\ but actually was considered
(implicitly) in \ihiggs). The interested reader should consult
\kirwan\ for more mathematically sound explanations of the contour
deformations arising in the similar context in the study of
symplectic quotients.

Perhaps a more illuminating way of understanding the contour
integral \cntr\ proceeds via the Duistermaat~-~Heckman formula
\DHf: \eqn\dhfr{\int_{X} {{\o}^n \over n!} e^{-{\m}[{\xi}]}  =
\sum_{f: V_{\xi} (f) = 0} {{e^{-
{\m}[{\xi}](f)}}\over{\prod_{i=1}^{n} w_i [{\xi}] (f)}}} Here
$(X^{2n},{\o})$ is a symplectic manifold (in the original DH setup
it should be compact, but the formula hold in more general
situation which extends to our case) with a Hamiltonian action of
a torus ${\bf T}^r$, ${\m}: X \to {\bf t}^*$ is the moment map,
${\xi} \in {\bf t} = Lie({\bf T})$ is the generator,  $V_{\xi} \in
Vect (X)$ is the vector field on $X$ representing the $\bf T$
action, generated by $\xi$, $f \in X$ runs through the set of
fixed points, and $w_i[{\xi}] (f)$ are the weights of the ${\bf
T}$ action on the tangent space to the fixed point.

In our case $X = {\widetilde{\CM}_{k}}$ is the moduli space of
instantons of charge $k$, ${\bf T}$ is the product of the Cartan
torus of $G$ and the torus ${\bf T}^2 \subset SO(4)$, ${\xi} = (a,
{\e}_1, {\e}_2)$ . The fixed points $f$ will be described in the
next section. However, already without performing the detailed
analysis of fixed points one can understand the meaning of \cntr:

\item{} {\ninepoint Suppose $f$ is a fixed point. It corresponds to
some quadruple $(B_1, B_2, I, J)$ such that the ${\bf
T}$-transformed quadruple belongs to  the same $U(k)$-orbit.
Working infinitesimally we derive that there must exist ${\phi}
\in Lie (U(k))$, such that: \eqn\undo{\eqalign{ & [ B_1, {\phi}] =
{\e}_1 B_1, \qquad [ B_2, {\phi}] = {\e}_2 B_2 \cr & - {\phi} I +
I a = 0, \qquad - a J + J {\phi} = - ({\e}_1 + {\e}_2) J \cr}} We
can go to the bases in the spaces $V$, $W$ where ${\phi}$ and $a$
are diagonal. Then the equations \undo\ will read as follows:
\eqn\undod{\eqalign{& \left( {\phi}_{I} - {\phi}_{J} + {\e}_1
\right) B_{1, IJ} = 0 \cr & \left( {\phi}_{I} - {\phi}_{J} +
{\e}_2 \right) B_{2, IJ} = 0 \cr & \left( {\phi}_I - a_l \right)
I_{I,l} = 0 \cr & \left( {\phi}_I +{\e}_1 +{\e}_2 - a_l \right)
J_{l,I} = 0 \cr}} For \undod\ to have a non-trivial solution some
of the combinations \eqn\cmbn{{\phi}_{IJ} +{\e}_{1,2}, {\phi}_I-
a_l, {\phi}_I +{\e}_1 + {\e}_2 - a_l} must vanish. This is where
the poles of the integrand \cntr\ are located. It follows from the
remark below \dfm\ that the equations \undod\ specify $\phi$
uniquely, given $a, {\e}_1, {\e}_2$ and $f$. Thus, exactly $k$ out
of $2k^2 + 2 kN$ combinations should vanish. Now let us look at
the (holomorphic) tangent space $T_{f} {\widetilde{\CM}_k}$. It is
spanned by the quadruples $({\d}B_1, {\d}B_2, {\d}I, {\d}J)$
obeying the linearized ADHM equations, and considered up to the
linearized $U(k)$ transformations, that is, it can be viewed as
first cohomology group of the following ``complex'':
\eqn\cmplext{C^0 = End(V) \longrightarrow C^1 = End(V) \otimes L
\oplus Hom (V, W) \oplus Hom(W, V)
 \longrightarrow C^2 = End(V) \otimes {\Lambda}^2 L}where the first arrow
 is given by the infinitesimal gauge transformation,
 while the second is the linearized ADHM equation: $d{\m}_c$.  To calculate
$\prod_i w_i [a, {\e}_1, {\e}_2](f)$ it is convenient to compute
first the ``Chern'' character:
\eqn\char{Ch(T_f{\widetilde{\CM}_k})  = \sum_{i} e^{w_i[a, {\e}_1,
{\e}_2] (f)}}which is given by the alternating sum of the Chern
characters of the terms in \cmplext: \eqn\euler{Ch (C^1) - Ch(C^0)
- Ch(C^2) = \sum_{I,J} e^{{\phi}_{I,J}} \left( e^{{\e}_1} - 1
\right) \left( 1 - e^{{\e}_2} \right) + \sum_{I, l} \left(
e^{{\phi}_I - a_l} + e^{a_l - {\phi}_I -{\e}} \right) } Upon the
standard conversion $$\sum_{\a} n_{\a} e^{x_{\a}} \mapsto
\prod_{\a} x_{\a}^{n_{\a}}$$ we arrive at \cntr. It remains to
explain why not every possible $k$-tuple out of the combinations
\cmbn\ contribute, only those which are picked by the $+i0$
prescription. This will be done in \promise. }

 \subsec{Classification of the residues}

The poles which with non-vanishing contributions to the integral
must have ${\phi}_{IJ} \neq 0$, for $I \neq J$, otherwise the
numerator vanishes. This observation simplifies the classification
of the poles. They are labelled as follows. Let $k = k_1 + k_2 +
\ldots + k_N$ be a partition of the instanton charge in $N$
summands which have to be non-negative (but may vanish), $k_l \geq
0$. In turn, for all $l$ such that  $k_l > 0$ let $Y_l$ denote a
partition of $k_l$: $$ k_l = k_{l,1} + \ldots k_{l, {\n}^{l,_1}},
\qquad k_{l,1} \geq k_{l,2} \geq \ldots \geq k_{l, {\n}^{l,1}} >
0$$ Let ${\n}^{l,1} \geq {\n}^{l,2} \geq \ldots
{\n}^{l,k_{l,1}}>0$ denote  the dual partition. Pictorially one
represents these partitions by the Young diagram with $k_{l,1}$
rows of the lengths ${\n}^{l,1}, \ldots {\n}^{l,k_{l,1}}$. This
diagram has ${\n}^{l,1}$ columns of the lengths $k_{l,1}, \ldots,
k_{l,{\n}^{l,1}}$. Sometimes we also use the notation $n_l =
{\n}^{l,1}$, and we find it useful to extend the sequence
$k_{l,i}$ all the way to infinity by zeroes: ${\bf k}_l = \{
k_{l,1} \geq k_{l,2} \geq \ldots k_{l,n_{l}+1} = k_{l, n_{l}+2} =
\ldots = 0\}$.

In total we have $k$ boxes distributed among $N$ Young tableaux
(some of which could be empty, i.e. contain zero boxes). Let us
label these boxes somehow (the ordering is not important as it is
cancelled in the end by the factor $k!$ in \cntr). Let us denote
the collection of $N$ Young diagrams by ${\vec Y} = (Y_1, \ldots,
Y_N)$. We denote by $\vert Y_l \vert = k_l$ the number of boxes in
the $l$'th diagram, and by $\vert \vec Y \vert = \sum_l \vert Y_l
\vert = \vert {\bf k} \vert = k$.

Then the pole of the integral \cntr\ corresponding to $\vec Y$ is
at ${\phi}_I$ with $I$ corresponding to the box $({\a},{\b})$ in
the $l$'th Young tableau (so that $0 \leq {\a} \leq {\n}^{l,{\b}},
\ 0 \leq {\b} \leq k_{l,{\a}}$) equal to: \eqn\pole{{\vec Y}
\longrightarrow {\phi}_{I} = a_l + {\e}_1 ({\a}-1) + {\e}_2 (
{\b}-1)}

\subsec{Residues and fixed points}

The poles in the integral \cntr\ correspond to the fixed points of
the action of the groups $G \times {\bf T}^2$ on the resolved
moduli space ${\tilde\CM}_{k,N}$. Physically they correspond to
the $U(N)$ (noncommutative) instantons which split as a sum of
$U(1)$ noncommutative instantons corresponding to $N$ commuting
$U(1)$ subgroups of $U(N)$. The instanton charge $k_l$ is the
charge of the $U(1)$ instanton in the $l$'th subgroup. Moreover,
these abelian instantons are of special nature -- they are fixed
by the group of space rotations. If they were commutative (and
therefore point-like) they had to sit on top of each other, and
the space of such point-like configurations would have been rather
singular. Fortunately, upon the noncommutative deformation the
singularities are resolved. The instantons cannot sit quite on top
of each other. Instead, they try to get as close to each other as
the uncertainty principle lets them. The resulting abelian
configurations were classified (in the language of torsion free
sheaves) by H.~Nakajima \nakajima.

Now let us fix a configuration ${\vec Y}$ and consider the
corresponding contribution to the integral over instanton moduli.
It is given by the residue of the integral \cntr\ corresponding to
\pole:

\eqn\res{\eqalign{&R_{\vec Y} =  {1\over{({\e}_1 {\e}_2)^k}}
\prod_{l} \prod_{{\a}=1}^{{\n}^{l,1}} \prod_{{\b}=1}^{k_{l,{\a}}}
{{{\CS}_{l}({\e}_1 ({\a}-1) + {\e}_2 ({\b}-1))}\over{({\e}
({\ell}(s)+1) - {\e}_2 h(s))({\e}_2 h(s) - {\e} {\ell} (s))}}
\times \cr & \prod_{l < m} \prod_{{\a}=1}^{{\n}^{l,1}}
\prod_{{\b}=1}^{k_{m,1}} \left( {{\left( a_{lm} + {\e}_1 ({\a} -
{\n}^{m,{\b}})+{\e}_2 (1-{\b}) \right) \left(  a_{lm} + {\e}_1
{\a} + {\e}_2 ( k_{l,\a} + 1 - {\b}) \right)}\over{\left( a_{lm} +
{\e}_1 {\a} + {\e}_2 ( 1 - {\b}) \right) \left( a_{lm} + {\e}_1
({\a} - {\n}^{m,{\b}}) + {\e}_2 ( k_{l,{\a}} + 1 - {\b})
\right)}}\right)^2 \cr}} where we have used the following
notations: $a_{lm} = a_l - a_m$,  \eqn\sss{{\CS}_{l}(x) = {Q(a_l +
x) \over { \prod_{m \neq l} ( x + a_{lm} ) (x + {\e} + a_{lm} )}
}, \qquad S_{l}(x) = {Q(a_l + x)\over {\prod_{m \neq l} ( x +
a_{lm})^2}} , } and \eqn\hkl{{\ell}(s) = {k}_{l,{\a}} - {\b},
\qquad h(s) = {k}_{l,{\a}} + {\n}^{l,{\b}} - {\a} - {\b} +1} Now,
if we set ${\e}_1 = {\hbar} = - {\e}_2$ the formula \res\ can be
further simplified. After some reshuffling of the factors it
becomes exactly the formula for the term in the sum \explctm,
corresponding to the partition $\{ k_{l,i} \}$.

\ndt {\bf Remark.} The expressions \res\explct\explctm\ are the
typical localization formulae for the instanton integrals. They
commonly appear in the two dimensional sigma model instanton
calculations, on the so-called A side. The Seiberg-Witten
prepotential \sw\ is the typical type B expression. In is not easy
to recognize in the type A expression the mirror manifold, and its
periods. To illustrate this point, we suggest to look at the
generating function of the number of holomorphic curves of genus
zero in the Calabi-Yau quintic, computed using localization
\maxim. It requires some extra work to map it to the mirror
calculation, yet it can be done \givental. In this paper we shall
not complete the story in the sense that we shall not prove
directly that our ``type A'' expression can be computed on the ``B
side'' involving Seiberg-Witten curves. We shall, however, present
a conjecture, which connects our calculation to its mirror
counterpart (that is, we shall define the mirror computation).

We shall also make some explicit checks for low instanton numbers
(up to five) to make sure we have computed the right thing.

\subsec{The first three nonabelian instantons}

We shall now give the formulae for the first three instanton
contributions to the prepotential for the general $SU(N)$ case,
with $N_f < 2N$.

We shall work with ${\e}_1 = {\hbar} = - {\e}_2$. It will be
sufficient to derive the gauge theory prepotential.

Directly applying the rules \cntr\res\ we arrive at the following
expressions for the moduli integrals: \eqn\zeds{\eqalign{& Z_1 =
{1\over{{\e}_1 {\e}_2}} \sum_{l} S_{l}  \cr & Z_2 = {1\over{(
{\e}_1 {\e}_2)^2}} \left( {1\over 4} \sum_{l} S_{l} \left( S_l (
+\hbar) + S_{l}( - \hbar) \right) + {1\over 2} \sum_{l \neq m}
{S_{l} S_{m} \over \left( 1 - {{\hbar}^2 \over a_{lm}^2}
\right)^{2} }\right)  \cr & Z_{3} = {1\over{({\e}_1 {\e}_2)^3}}
\Biggl( \sum_{l} {S_{l}  \left( S_{l} (+ \hbar) S_{l}(+ 2 \hbar) +
S_{l}( - \hbar) S_{l} (- 2\hbar) + 4 S_{l}( + \hbar) S_{l}(-
\hbar) \right)\over 36} + \cr  & \qquad\qquad\qquad\qquad \sum_{l
\neq m} {S_{l} S_{m} \over 4\left( 1 - { 4{\hbar}^2 \over
a_{lm}^2} \right)} \Biggl( S_{l} (+ \hbar ) \left( 1 - {2
(\hbar/a_{lm})^2 \over  ( 1 - (\hbar/a_{lm}) )} \right)^{2} + \cr
& \qquad\qquad\qquad\qquad\qquad\qquad\qquad\qquad S_{l}(- \hbar)
 \left( 1 - {2 (\hbar/a_{lm})^2 \over ( 1 + (\hbar/a_{lm})))}
\right)^{2} \Biggr) + \cr & \qquad\qquad\qquad\qquad\qquad \sum_{l
\neq m \neq n} {S_{l} S_{m} S_{n} \over 6 \left( \left( 1 -
{\hbar^2 \over a_{lm}^2 } \right)\left( 1 - {\hbar^2 \over
a_{ln}^2 } \right)\left( 1 - {\hbar^2 \over a_{mn}^2 } \right)
\right)^{2}} \Biggr)
 \cr & \cr
 &  {\rm where}\qquad S_l = S_l (0), \qquad S_{l}^{(n)} = {\p}_x^n S_l (x) \vert_{x=0} \ , \cr }}
which yield: \eqn\prep{\eqalign{& {\CF}_1 = \sum_{l} S_{l} \cr &
{\CF}_2 = \sum_{l} {1\over 4} S_{l} S_{l}^{(2)} + \sum_{ l \neq m}
{S_{l} S_{m}  \over a_{lm}^2}  + O({\hbar}^2)\cr & {\CF}_3 =
\sum_{l} {S_{l} \over 36} \left( S_{l} S_{l}^{(4)} + 2 S_{l}^{(1)}
S_{l}^{(3)} + 3 S_{l}^{(2)} S_{l}^{(2)} \right) + \cr & \qquad
\sum_{l \neq m} {S_{l} S_{m} \over a_{lm}^4}\left( 5 S_{l}  - 2
a_{lm} S_{l}^{(1)} + a_{lm}^2 S_{l}^{(2)} \right) + \cr & \qquad
\sum_{l\neq m\neq n} {2 S_{l} \ S_{m} S_{n} \over 3 ( a_{lm}
a_{ln} a_{mn})^2 } \left(a_{ln}^2 + a_{lm}^2 + a_{mn}^2 \right)  +
O ({\hbar}^2) \cr}}

\subsec{Four and five instantons}

To collect more ``experimental data-points'' we have considered in
more details the cases of the gauge groups $SU(2)$ and $SU(3)$
with fundamental matter. We have computed explicitly the
prepotential for four and five instantons and found  a perfect
agreement (yet a few typos) with the results of \experiment. We
should stress that this is a non-trivial check. Just as an
example, we quote here the expression for ${\CF}_5$ for $SU(2)$
gauge theory with $N_f = 3$:
$${\CF}_5 (a, m) = {{{\m}_3}\over{8 a^{18}}} ( 35 a^{12} - 210
a^{10} {\m}_2 + a^8 \left( 207 {\m}_2^2 + 846 {\m}_4 \right) $$ $$
- 1210 a^6 {\m}_2 {\m}_4 + a^4 \left( 1131 {\m}_4^2 + 3698
{\m}_3^2 {\m}_2 \right) - 5250 a^2 {\m}_3^2 {\m}_4 + 4471 {\m}_3^4
),$$ where $2a = a_1 - a_2$, ${\m}_2 = m_1^2 + m_2^2 + m_3^2$,
${\m}_3 = m_1 m_2 m_3$, ${\m}_4 = (m_1 m_2)^2 + (m_2 m_3)^2 + (m_1
m_3)^2$.

\subsec{Adjoint matter and other matters}

So far we were discussing ${\CN}=2$ gauge theories with matter in
the fundamental representations. Now we shall pass to other
representations. It is simpler to start with the adjoint
representation. The ${\e}$-integrals \cntr\ reflect  both the
topology of the moduli space of instantons and also of the matter
bundle.

The latter is the bundle of the Dirac zero modes in the
representation of interest. For the adjoint representation, and on
${\IR}^4$, this bundle can be identified with the tangent bundle
to the moduli space of instantons. Turning on a mass term for the
adjoint hypermultiplet corresponds to working equivariantly with
respect to a certain $U(1)$ subgroup of the extended R-symmetry
group.  The $U(1) \times G \times {\bf T}^2$ equivariant Euler
class of the tangent bundle (= the $G \times {\bf T}^2$
equivariant Chern polynomial) is the instanton measure in the case
of massive adjoint matter. This reasoning leads to the following
${\e}$-integral: \eqn\adjcntr{\eqalign{& Z_{k} = {1 \over k!}
\left( {({\e}_1 + {\e}_2)( {\e}_1 + m ) ({\e}_2 + m) \over 2{\pi}
i \ {\e}_1 {\e}_2 \ m \ ( {\e} - m)} \right)^{k} \oint
{\prod_{I=1}^{k} }{{{\rm d}{\phi}_I} \ P ({\phi}_I + m)  P (
{\phi}_I + {\e} -m) \over \ \ \ P ({\phi}_I ) P ({\phi}_I + {\e}
)} \times \cr & \qquad \qquad \qquad  \qquad \qquad \qquad \times
\prod_{I<J} {{\phi}_{IJ}^2 ({\phi}_{IJ}^2 - {\e}^2) ({\phi}_{IJ}^2
- ( {\e}_1 - m)^2)({\phi}_{IJ}^2 - ({\e}_2 - m)^2) \over
({\phi}_{IJ}^2 - {\e}_1^2 )({\phi}_{IJ}^2 -
{\e}_2^2)({\phi}_{IJ}^2 - m^2 )({\phi}_{IJ}^2 - ({\e}-m)^2)} \cr}}
Note the similarity of this expression to the contour integrals
appearing\dbound\ in the calculations of the bulk contribution to
the index of the supersymmetric quantum mechanics with $16$
supercharges (similarly, \cntr\ is related to the one with $8$
supercharges). This is not an accident, of course.

\ndt Proceeding analogously to the pure gauge theory case we
arrive at the following expressions for the first two instanton
contributions to the prepotential (which agrees with \experiment):
\eqn\adjprep{\eqalign{& {\CF}_1 = m^2 \sum_{l} T_{l} \cr & {\CF}_2
= \sum_l \left( - {3m^2 \over 2} T_l^2 + {m^4 \over 4} T_l
T_{l}^{(2)} \right) + m^4 \sum_{l \neq n} T_l T_n \left( {1\over
a_{ln}^2} - {1\over 2 (a_{ln} + m)^2} - {1\over 2(a_{ln} - m)^2}
\right) \cr}} where $T_l (x) = \prod_{n \neq l} \left( 1 - {m^2
\over (x + a_{ln})^2} \right)$, $T_l = T_l(0)$, $T_l^{(n)} =
{\p}_x^n T_l(x) \vert_{x=0}$ (cf. \hollowood).
\subsec{Perturbative part}

So far we were calculating the nonperturbative part of the
prepotential. It would be nice to see the perturbative part
somewhere in our setup, so as to combine the whole expression into
something nice.

One way is to calculate carefully the equivariant Chern character
of the tangent bundle to ${\widetilde{\CM}_k}$ along the lines
sketched in the end of the previous section\promise. The faster
way in the ${\e}_1 + {\e}_2 = 0$ case is to note that the
expression \explct\ is a sum over partition with the universal
denominator, which is not well-defined without the non-universal
numerator. Nevertheless, let us try to pull it out of the sum.

We get the infinite product (up to an irrelevant constant): $$
\prod_{i,j = 1}^{\infty} \prod_{l \neq  n} {1\over{a_{ln} +
{\hbar} (i - j)}} \sim $$ \eqn\schwinger{ {\exp} - \sum_{l \neq n}
\int_{0}^{\infty} {{ds} \over s} {e^{- s a_{ln}} \over{(e^{\hbar
s}- 1) ( e^{-\hbar s} -1)}}} If we regularize this by cutting the
integral at $s \sim {\ve} \to 0$, we get a finite expression,
which actually has the form $${\exp} {{\CF}^{pert} ( a , {\e}_1,
{\e}_2) \over{{\e}_1 {\e}_2}}, $$ with ${\CF}^{pert}$ being
analytic in ${\e}_1, {\e}_2$ at zero. In fact $${\CF}^{pert} (a,
0, 0) = \sum_{l \neq n} {1\over 2} a_{ln}^2 {\rm log} \ a_{ln} +
{\rm  \ ambiguous \ quadratic \ polynomial \ in} \ a_{ln} . $$
 The formula \schwinger\ is a
familiar expression for the Schwinger amplitude of a mass $a_{ln}$
particle in the electromagnetic field \eqn\emf{F \propto {\e}_1 \
dx^1 \wedge dx^2 + {\e}_2 \ dx^3 \wedge dx^4 \ .} Its appearance
will be explained in the next section.

\ndt Let us now combine ${\CF}^{inst}$ and ${\CF}^{pert}$ into a
single ${\e}$-deformed prepotential $$ {\CF} (a, {\e}_1, {\e}_2) =
{\CF}^{pert} (a, {\e}_1, {\e}_2 ) + {\CF}^{inst} (a, {\e}_1,
{\e}_2)$$ where for  general ${\e}_1, {\e}_2$ we define:
\eqn\schw{{\CF}^{pert} (a, {\e}_1, {\e}_2 ) = \sum_{l \neq n}
\int_{{\ve}}^{\infty} {{ds \over s}} {e^{- s a_{ln} } \over  {\rm
sinh} \left( {s {\e}_1 \over 2} \right) {\rm sinh} \left( {s
{\e}_2 \over 2 } \right) }} with the singular in ${\ve}$ part
dropped. We define: \eqn\prtn{{\CZ} ( a, {\e}_1, {\e}_2 ; q) =
{\exp} {{\CF} (a, {\e}_1, {\e}_2 ; q) \over {{\e}_1 {\e}_2}} }

\newsec{M- and K-theory inspired conjectures}

In this section we suggest a physical interpretation to the
${\e}$-deformed prepotential ${\CF} (a,{\e}_1, {\e}_2; q)$. We
also conjecture a relation of ${\CZ} (a, {\hbar}, -{\hbar};q )$ to
a tau-function of Toda hierarchy.

\ndt {\bf Philosophy.} So far we studied the equivariant
cohomology of the instanton moduli space, pushforwards, and
localization. The equivariant cohomology is a quasiclassical limit
of the equivariant K-theory (in the same sense in which the DH
formula is the quasiclassical limit of the (super)character
formula for ${\Tr} (-)^F e^{- {\hat\m} [{\xi}]}$. Conversely, by
studying the ${\bf S}^1$-equivariant cohomology of the loop space
of the original space one can get the index theorems natural in
K-theory \niemi.

This is of course an old story. However, this old story might get
a new meaning with the entering of M-theory on the scene.

\subsec{Five dimensional viewpoint}

It was understood long time ago that the instanton corrections to
the prepotential of ${\CN}=2$ gauge theory can be interpreted as
one-loop corrections in the five dimensional theory compactified
on a circle, in the limit of vanishing circle radius \nikfive.

Consider five dimensional ${\CN}=2$ gauge theory with the gauge
group $G$, and possibly some matter. Consider a path integral in
this theory on the space-time manifold which is a product ${\bf
S}^1 \times {\IR}^4$. We shall impose periodic boundary conditions
on the fermions in the theory (up to a twist described
momentarily). We shall also consider the vacuum in which the
adjoint scalar in the vector multiples has vacuum expectation
value ${\vf} \in {\bf t}$. We should also specify the holonomy $g
\in T $ of the gauge field around the circle at infinity of
${\IR}^4$ (which must commute with ${\vf}$). Together they define
an element ${\bf a} = g {\exp} {\b}{\vf} \sim {\exp} {\b} a$ of
the complexified Cartan {\it subgroup} ${\bf T}_{\bf C}$ of the
gauge group \nikfive.

Let us define the following generalized index: \eqn\gnind{{\CZ}
({\bf a}, {\e}_1, {\e}_2, {\e}_3; {\b}) = {\Tr}_{{\CH}_{\bf a}}
(-)^{2(j_L + j_R)} {\exp} - \left[ \left( {\e}_1 - {\e}_2 \right)
J_L^3 + \left( {\e}_1 + {\e}_2 \right) J_R^3 + {\e}_3 J_I^3 + {\b}
H \right]} We now choose ${\e}_3 = {\e}_1 + {\e}_2$. This is the
counterpart of choosing the subgroup $SU(2)_d$ as we did in the
section $2$. Here we have used the little group $SO(4)$ spins
$j_L, j_R$ and the generator $J_I^3$ of the R-symmetry group
$SU(2)_I$. With this choice of ${\e}$'s some of the supercharges
of the five dimensional gauge theory will commute with the twists
$e^{{\e} \cdot J}$ and \gnind\ will define a generalized index.

The five dimensional theory has two kinds of particles. The
perturbative spectrum consists of the gauge bosons and their
superpartners (we shall now consider the case of minimal susy
theory for simplicity). In addition, the theory has solitons,
coming from instanton solutions of four dimensional gauge theory.
To find the spectrum of these solitons and their bound states one
can adopt the standard collective coordinate quantization scheme.

In the limit ${\b} \to 0$ the infinite-dimensional version of the
heat kernel expansion will reduce the supertrace in \gnind\ to a
path integral in the four dimensional gauge theory. Moreover, the
arrangement of the twists is such that the theory will possess
some supersymmetry, which we identify with ${\tilde Q}$. One can
play with the gauge coupling to further reduce the path integral
to a finite-dimensional integral over the instanton moduli space,
of the kind we considered in this paper.

On the other hand, geometrically, the twists in \gnind\ can be
realized by replacing the flat five dimensional space-time by a
twisted ${\IR}^4$ bundle over the circle ${\bf S}^1$ of the radius
${\b}/2{\pi}$ such that by going around this circle one twists
the\foot{If ${\e}_1 + {\e}_2 \neq 0$ one should also twist the
transverse six dimensional space} ${\IR}^4$ according to \gnind.
Let us denote the resulting (locally flat) space by $X_{\e}$.

Now imagine engineering \vafaengine\ the five dimensional gauge
theory by ``compactifying'' M-theory on a non-compact Calabi-Yau
given by the appropriate fibration of the ALE singularity over the
base ${\IP}^1$. By further compactifying on $X_{\e}$ we get a
background, which can be now analyzed string-theoretically.

If we view the base circle of $X_{\e}$ as the $M$-theory circle,
then we end up with the IIA Mellin-like background, where one has
a vev of the RR 1-form field strength, which is actually
\eqn\graviph{ F = {\e}_1 dx^1 \wedge dx^2 + {\e}_2 dx^3 \wedge
dx^4} near $x=0$. In fact, by going to the weak string coupling
limit (${\b} \to 0$) one can make \graviph\ to hold arbitrarily
far away from the origin.

The five dimensional particles, going around the circle ${\bf
S}^1$ appear in the four remaining dimensions like particles
carrying some charge with respect to $F$. In calculating their
contribution to the supertrace \gnind\ we would perform the
standard Schwinger calculation, as in \gopakumarvafa.

This identification suggests the interpretation of $\hbar$. Let us
expand ${\CF} ( a, {\hbar}, - {\hbar})$ as a power series in
${\hbar}$: \eqn\genera{-{\rm log} {\CZ} (a, {\hbar}, - {\hbar}) =
{1\over{{\hbar}^2}} {\CF} (a , {\hbar}, - {\hbar}) = \sum_{g =
0}^{\infty} {\hbar}^{2g-2} {\CF}_{g} (a)} (the fact that only the
even powers of ${\hbar}$ appear follows from the obvious symmetry
${\e}_1 \leftrightarrow {\e}_2$). The expansion \genera\ suggests
that ${\hbar}$ has to play a role of the string coupling constant.
To be more precise, in the setup in which the gauge theory is
realized as IIA compactification on the ALE singularity fibered
over ${\IP}^1$ the prepotential is essentially calculated by the
worldsheet instantons of genus zero. The higher genus corrections
give rise to the $R^2 F^{2g-2}$ couplings \gravilit, where $R$ is
the curvature of the four dimensional metric, and $F$ is the
graviphoton field strength. Collecting all the evidence above we
conjecture that the ${\e}$-deformed prepotential captures the
graviphoton couplings (even in the case ${\e}_1 + {\e}_2 \neq 0$,
where the graviphoton field strength is not self-dual -- in this
case the theory should be properly twisted).

\subsec{K-theory viewpoint}

If we don't take the limit ${\b} \to 0$ we can still give a finite
dimensional expression for \explct. The collective coordinate
quantization leads \nikfive\ to the minimal supersymmetric quantum
mechanics on $\widetilde{{\CM}_k}$, whose ground states correspond
to the harmonic spinors on $\widetilde{{\CM}_k}$. The index
\gnind\ calculates the equivariant index of Dirac operator on
$\widetilde{{\CM}_k}$. The latter has the following Atiyah-Singer
expression: \eqn\atsng{{\CZ} \left( {\bf a}, {\e}_1, {\e}_2 , {\b}
; q \right) = \sum_{k=0}^{\infty} q^k \oint_{\widetilde{{\CM}_k}}
\widehat{A_{\b}} \left( \widetilde{{\CM}_k} \right) }where ${\hat
A}(M)$ is the A-roof genus of the manifold $M$: \eqn\aroof{{\rm
ch}(TM) = \sum_i e^{x_i} \qquad \Longrightarrow \qquad {\hat
A}_{\b} (M) = \prod_i {{\b} x_i \over e^{{\b}x_i \over 2} -
e^{-{{\b}x_i \over 2}}}} The localization technique (this time in
equivariant K-theory \atiyahsegal) leads to the following
expression for \atsng: \eqn\atsngii{{\CZ} \left( {\bf a}, {\hbar},
-{\hbar}, 2{\b} ; q \right) = \sum_{{\vec{\bf k}}} q^{\vert
\vec{\bf k} \vert} \prod_{(l,i) \neq (n,j)} {{{\rm sinh} \ {\b}
\left( a_{ln} + {\hbar} \left( k_{l,i} - k_{n,j} + j - i
\right)\right)} \over {{\rm sinh} \ {\b} \left( a_{ln} + {\hbar}
\left(  j - i \right)\right)}}}

It would be nice to analyze \atsngii\ further, relate it to the
relativistic Toda chain spectral curves \nikfive, and to the four
dimensional analogue of Verlinde formula \avatars. It also seems
reasonable to expect applications of \atsngii\ to the DLCQ
quantization of the M5-brane \abkss.

\subsec{Chiral fermions and M5 brane}

Another conjecture relates the expansion \explct\ to the dynamics
of the Seiberg-Witten curve\foot{For simplicity we consider the
case $N_f=0$. The conjecture for $N_f > 0$ case is easy to
guess.}. Denote, as before $q = {\Lambda}^{2N}$.

\ndt Consider the theory of a free complex chiral fermion ${\psi},
{\psi}^*$,  \eqn\frferm{{\CS} = \int_{\Sigma} {\psi}^* {\pb}
{\psi}} living on the curve $\Sigma$: \eqn\swcurv{w +
{{\Lambda}^{2N} \over w} ={\bf P} ({\l}), \qquad {\bf P} ({\l}) =
\prod_{l=1}^{N} ( {\l} - {\a}_l )} embedded into the space ${\IC}
\times {\IC}^*$ with the coordinates $({\l}, w)$. This curve has
two distinguished points $w = 0$ and $w = \infty$ which play a
prominent role in the Toda integrable hierarchy \todalit. Let us
cut out small disks $D_{0}$ and $D_{\infty}$ around these two
points.

\ndt  The path integral on the surface ${\Sigma}$ with two disks
deleted will give a state in the tensor product  ${\CH}_0 \otimes
{\CH}^{*}_{\infty}$ of the Hilbert spaces ${\CH}_0$,
${\CH}_{\infty}$ associated to ${\p}D_0$ and ${\p}D_{\infty}$
respectively. It can also be viewed as an operator $G_{\Sigma} :
{\CH}_0 \to {\CH}_{\infty}$.

\ndt  Choose a vacuum state $\vert 0 \rangle \in {\CH}_0$ and its
dual $\langle 0 \vert \in {\CH}_{\infty}^*$ (we use the global
coordinate $w$ to identify ${\CH}_0$ and ${\CH}_{\infty}$).
Consider \eqn\tu{{\tau}_{\Sigma}   = \left\langle 0 \left\vert
{\exp} \left( {1\over {\hbar}} \oint_{{\p}D_\infty} \ S \ J
\right) \ G_{\Sigma} \ {\exp} \left( - {1\over{\hbar}}
\oint_{{\p}D_{0}} S \ J \right) \right\vert 0 \right\rangle}
where: \eqn\not{\eqalign{&  J = : {\psi}^* {\psi} : \cr & dS =
{1\over{2\pi i}} {\l} {dw \over w} \cr}}and  we choose the branch
of $S$ near $w = 0, \infty$ such that (cf. \whitham) : $$S =
{N\over{2\pi i}} {w}^{\mp {1\over N}} + O ( {\l}^{-1} ) $$ Let us
represent $\Sigma$ as a two-fold covering of the ${\l}$-plane. It
has branch points at ${\l} = {\a}^{\pm}_l$ where $${\bf P} (
{\a}^{\pm}_l ) = \pm 2 \Lambda^{N}$$ Let us choose the cycles
$A_l$ to encircle the cuts between ${\a}_l^{-}$ and
${\a}_{l}^{+}$. Of course, in $H_1 ({\Sigma}, {\IZ})$, $\sum_l A_l
= 0$. Then, we define: $$a_l =  \oint_{A_l} dS \ .$$Our final
conjecture states: \eqn\fnc{\mathboxit{{\CZ} (a, {\hbar}, -
{\hbar}; q) = {\tau}_{\Sigma} }} Note that from this conjecture
the fact that ${\CF}_{0} (a, 0,0) $ coincides with the
Seiberg-Witten expression follows as a consequence of the
Krichever universal formula \kricheverwhitham. The remaining
paragraph is devoted to the explanation of the motivation behind
\fnc.

Let us assume that we are in the domain where ${\a}_l - {\a}_m \gg
{\Lambda}$. Then the surface ${\Sigma}$ can be decomposed into two
halves ${\Sigma}_{\pm}$ by $N$ smooth circles $C_l$ which are the
lifts to $\Sigma$ of the cuts connecting ${\a}_l^{-}$ and
${\a}_l^{+}$. The path integral calculating the matrix element
\tu\ can be evaluated by the cutting and sewing along the $C_l$'s.
The path integral on ${\Sigma}_{\pm}$ gives a state in
$$\otimes_{l=1}^{N} {\CH}_{C_l}$$ (its dual). If we first pull the
$\oint S J$ as close to $C_l$ as possible, we shall get the
Hilbert space obtained by quantization of the fermions which have
$a_l + {\half} mod {\IZ}$ moding: \eqn\psin{{\psi} (w) \sim
\sum_{i \in {\IZ}} {\psi}_{l,i} w^{a_l + i} \left( {dw \over  w}
\right)^{\half}}near $C_l \subset \Sigma$. In addition, the states
in ${\CH}_{C_l}$ of fixed total $U(1)$ charge are labelled by the
partitions $k_{l,i}$. We conjecture, that \eqn\fncc{\left\langle 0
\left\vert  e^{\oint S J} \prod_{l,i} {\psi}_{l,k_{l,i} - i}
{\psi}^{*}_{l,-i} \right\vert 0 \right\rangle_{l} \sim
\prod_{(l,i) < (m,j)} ( a_{lm} + {\hbar} ( k_{l,i} - k_{m,j} + j -
i) )} It is clear that \fncc\ implies \fnc. For $N=1$ \fncc\ is of
course a well-known fact (with the coefficient given by $\prod_{i
< j} {1\over{j-i}}$), which leads to the following formula for
\explct\foot{which can also be derived using Schur identities
\macdonald}:
$$
Z_{N=1}({\hbar, - \hbar}; q) = e^{- {q\over{{\hbar}^2}}}
$$confirming the fact, that even though we worked with the
resolved moduli space $\cup_k \widetilde{\CM}_{k,1} =\cup_k
({\IC}^2)^{[k]}$ the ``symplectic'' volume we got is that of
$\cup_k {\widetilde M}_{k,1} = \cup_k {Sym}^k ({\IR}^4)$.

The conjecture \fnc\ should in some sense trivially follow from
the M5/NS5 realization of the four dimensional supersymmetric
gauge theory\wittensolution\warner. The ${\e}$-twisting of the
four dimensional part of the fivebrane worldvolume reduces the
dynamical degrees of freedom down to those of the chiral boson on
the curve, which after fermionization should lead to \fnc.

Finally, if \fnc\ is true, it is natural to conjecture that the
analogous equivariant generating functions for instantons on ALE
spaces of the ADE type will be related to the ADE WZW theories on
the SW curves.

\footatend\vfill\supereject\immediate\closeout\rfile\writestoppt
\baselineskip=14pt\centerline{{\bf References}}\bigskip{\frenchspacing%
\parindent=20pt\escapechar=` \input refs.tmp\vfill\eject}\nonfrenchspacing \bye